\documentclass[12pt]{article}

\input epsf
\def\beq{\begin{eqnarray}}
\def\eeq{\end{eqnarray}}

\def\e{{\varepsilon}}
\def\t{{\theta}}

\def\g{gluodynamics}

\def\lsim{\mathrel{\rlap{\lower3pt\hbox{\hskip0pt$\sim$}}
    \raise1pt\hbox{$<$}}}         
\def\gsim{\mathrel{\rlap{\lower4pt\hbox{\hskip1pt$\sim$}}
    \raise1pt\hbox{$>$}}}         

\begin{document}

\begin{titlepage}
\thispagestyle{empty}
\begin{flushright}
TPI-MINN-02/19\\
UMN-TH-2104/02 \\
\end{flushright}

\vskip 1cm
\begin{center}
{\Large \bf QCD Vacuum and Axions: What's Happening?}

\vskip 1cm
{Gregory Gabadadze and M. Shifman}

\vskip 1cm
{\it Theoretical Physics Institute, University of Minnesota,
Minneapolis,
MN 55455}\\
\end{center}

\vspace{0.9cm}
\begin{center}
{\bf Abstract}
\end{center}

A deeper understanding of the vacuum structure in  QCD invites
one to rethink certain aspects of the axion physics.
The recent advances are mostly due to developments in supersymmetric
gauge theories and the brane theory, in
which QCD can be embedded.  They include,
but are not limited to, the studies of metastable
vacua in multicolor gluodynamics,  and the domain walls.
We   briefly review   basics of the axion physics
and then present a   modern  perspective on
a rich interplay between  the QCD vacuum structure and  axion physics.

\vspace{0.1in}

\vspace{2.5in}

{--------------------------------------- \\
An expanded version of a review talk given by G.G. at
{\it ``Continuous Advances in QCD 2002/Arkadyfest''},
honoring the 60'th birthday of Arkady Vainshtein, Minneapolis,
May 17 - May 23, 2002.}

\end{titlepage}

\newpage

\section{Introduction}
\vspace{0.5cm}

Almost 25 years elapsed since the axion was
introduced in particle physics \cite{Weinberg,Wilczek} as a
possible solution of  the strong $CP$ problem. Since then,
it became a
 text-book and encyclopedia subject. For instance,
 Oxford University's  {\em ``A Dictionary of Science''} defines axion
as ``a hypothetical elementary particle postulated to explain
why there is no observed
$CP$  violation (see $CP$ invariance) in the strong interaction
(see fundamental interactions). Axions have not been detected
experimentally, although it has been possible to put limits on
their mass and other properties from the effects that they would
have on some astrophysical phenomena (e.g. the cooling of stars).
It has also been suggested that they may account for some or all
of the missing mass in the universe."

While at the early stages the axion physics was considered
predominantly in the context of quantum chromodynamics, at
present the center of gravity of the axion studies shifted
in astrophysics. It was realized rather early that the axion
was a viable dark matter candidate.
The research on this aspect of the axion physics quickly picked
up and never subsided.
Extensive investigations were and are   being carried out in the
astrophysical community.
At the same time, after the rapid advances in the 1980's, the QCD
practitioners
seemingly lost interest in this subject. The reason is obvious:
the progress in understanding the QCD vacuum structure
was painfully slow. The prevailing  impression  was that
``nothing happened here,"
so there was no motivation for revisiting
QCD-related aspects of the axion physics.

In this review we will try to argue that ``something interesting
happened here."
A substantial progress has been achieved in the recent years
mainly due to  insights in QCD obtained from supersymmetry and the
brane theory.
The existence of a multitude of (quasi) stable vacua at large $N_c$
and ``abnormally" thin domain walls with ``abnormally" heavy
excitations are just a few topics of interest that should be mentioned
in this context.
A deeper understanding of the QCD vacuum structure   requires
a reassessment of a number of issues of direct relevance to axions.
After a brief summary of basics  of the axion physics we review these
new developments.

\section{The strong {\boldmath ${CP}$} problem}

\subsection{The {\boldmath $\theta$} term}

After the discovery of  asymptotic  freedom in QCD in 1973
\cite {AF1,AF2} for a short while
it was believed  that QCD  possesses the same natural conservation laws
as its more primitive predecessor, QED.
The discovery that $P$ and $T$ conservation in QCD is not natural
came as a shocking surprise. This fact was realized with the
advent of
 instantons \cite {Polyakov} which demonstrated that the so-called
$\theta$   term
\beq
\Delta \,{\cal L}_{\t}\,\equiv \,{\t\over 32\,\pi^2}\,G_{\mu\nu}^a\,
{\tilde G}_{\mu\nu}^a\,,
\label{Lt}
\eeq
does not necessarily vanish.
Here the dual field strength is defined as $$ {\tilde G}_{\mu\nu}^a\,
\equiv \, (1/2)\,\e_{\mu\nu\alpha\beta}\,G_{\alpha\beta}^a\,.$$
(The indices are assumed to be contracted via the flat space metric).
The operator $G\tilde G$ has dimension four, it  can
and should  be added to the QCD Lagrangian.
The $\theta$ term (\ref {Lt})
violates $P$ and $T$ invariance (and hence, it violates $CP$
since $CPT$ is preserved). Note that the
 analogous term $\Delta \,{\cal L}\sim F\tilde F$ in QED has no
impact on the theory whatsoever.
What is the difference?

The $\theta$  term can be rewritten as follows
\beq
\Delta \,{\cal L}_{\t}\,=\,\t \, \partial_{\mu}\,K_{\mu}\,,
\label{LK}
\eeq
where $K_\mu$ is  the Chern-Simons current defined as
\beq
K_{\mu}\,=\,{1 \over 16\,\pi^2}\,\e_{\mu\nu\alpha\beta}\,
\left (A_\nu^a\,\partial_{\alpha}\,A_{\beta}^a\,+\,{1\over 3}\,
f_{abc} A_\nu^a\,A_\alpha^b \,A_\beta^c  \right )\,.
\label{CS}
\eeq
Being a total derivative, the $\theta$  term does not  affect the
equations of motion. At a naive level, one can discard in the action
the integrals over full derivatives. This was a rationale behind
the original belief that QCD naturally conserves $P$ and $T$.

The instantons revealed the fact that the vacuum structure in QCD is
more complicated than that in QED.
In particular, the field configurations with the instanton
boundary conditions give rise to a nonvanishing
\beq
(\Delta S_\theta )_{\rm one~inst} = \int d^4x \,
(\Delta \,{\cal L}_{\t})_{\rm one~inst} =\theta \,.
\label{done}
\eeq
The integral over the full derivative does not vanish \footnote{
We jumped here from the Minkowski to the Euclidean formulation of the
theory. In passing from Minkowski to Euclidean, the $\theta$ term
(\ref {Lt})
acquires an ``i'' factor, so does the integration measure in the
action. Since this is a text-book topic,
we will pass freely from Minkowski to Euclidean and back
making no explicit statements as to which space any  given
formula belongs to.}.
Therefore,  $CP$-violating
effects may  be present in strong interactions.

We pause here to make an explanatory remark
regarding Eq.  (\ref{done}).
A key notion is the topological charge ${\cal V}$ of a gauge field configuration,
\begin{eqnarray}
{\cal V}&\equiv &\int\,d^4x\,\partial_{\mu}\,K_{\mu}\,=\,\int\,d^3x\,
K_0(x, t)|^{t=+\infty}_{t=-\infty}\nonumber\\[3mm]
&\equiv & {\cal K}(t=+\infty)- {\cal K}(t=-\infty)\,,
\label{Q}
\end{eqnarray}
where ${\cal K}$ is usually referred to as
the  Pontryagin number.
The topological charge is zero for any perturbative
gauge fields --- such fields are said to have trivial topology.
The instanton field configuration has a nontrivial topology.
In the $A_0=0$ gauge it interpolates between
 $A_{m}(x, t \to -\infty)\,=\,0\,,~m=,1,2,3,$
and
\beq
A_{m}(x, t \to +\infty)\,=\,U^+\,
\partial_m \,U\,,
\label{asym}
\eeq
where the matrix $U$ is Poyakov's hedgehog,
\beq
U(\vec x) =\exp\left( -\frac{i\pi\vec x\vec\sigma}{\sqrt{\vec x^2 +\rho^2}}\right).
\eeq
The Pontryagin number for $U$ reduces to
\beq
{\cal K}=\,{1 \over 24\,\pi^2  }\,\int\,d^3x\,\e_{ijk}\,{\rm Tr}
\,\left (U^+\,\partial^i\,U \right )\,\left (U^+\,\partial^j\,U \right )\,
\left (U^+\,\partial^k\,U \right )=1 \,,
\label{top}
\eeq
implying that the instanton topological charge is unity.

\subsection{Superselection rule and {\boldmath $\theta$}
 sectors}

As we pointed out in the previous section,
the value of the parameter $\t$ {\it a priori}
can be arbitrary. The theories  with different values of
$\t$ describe different worlds which do not ``communicate'' with
each other. In other words, the worlds with different $\t$
belong to distinct superselection sectors \cite{Jackiw,DGross}.

To see that this is the case and to further  elucidate the role of $\t$
let us consider pure gluodynamics in
the Hamilton gauge $A_0=0$. In this gauge the Lagrangian does not depend
on $A_0$ and the Gauss'  law (which in other gauges
could have been obtained by varying  the action with respect to $A_0$)
is   imposed as a constraint on physical states
\beq
D^i\,G_{i0}\,|{\rm Phys}\rangle \,=\,0\,.
\label{Gauss}
\eeq
This gauge fixing does not eliminate, however, the gauge
freedom completely. Purely spatial
gauge transformations (independent of the time variable)
are still allowed.
The generator of these residual gauge
transformations can be written as
\beq
{\cal G}(\alpha) \, \equiv \,{\rm exp}
\left (i\,\int\,d^3x\,{\rm Tr}\,D^i\,G_{i0}({\vec x})\,
\alpha({\vec x})\right ) \,,
\label{G}
\eeq
where $\alpha \,=\,\alpha^a\,t^a$ and the trace ${\rm Tr}$
runs over the color indices. The  generator ${\cal G}(\alpha) $ acts
on the spatial components of the gauge fields,
  \beq
{\cal G}^+\, A_k\,{\cal G}\,=\,U^+\,
\left ( A_k\,+\,i\,\partial_k \right)\,U\,,
\label{GA}
\eeq
where    $U \equiv {\rm exp}(i\alpha(x))$.
Furthermore, it is straightforward to show that the operator ${\cal G} $
does not commute with ${\cal V}$ if the corresponding
gauge transformations give rise to a nonzero right-hand side
in Eq. (\ref {top})
(the latter are called {\it large gauge transformations}),
\beq
[\,{\cal G}\,{\cal V}\,]\,\ne\,0\,.
\label{GQ}
\eeq
Therefore, an eigenstate of ${\cal V}$ cannot be a
physical state.
Instead, the physical state is defined as  a superposition of
the eigenstates of ${\cal V}$
\beq
|\t \rangle \,=\,\sum_{n=-\infty}^{+\infty}\,e^{i\,\t\,n}\,|n \rangle \,,
\label{tvac}
\eeq
where ${\cal V} |n \rangle \,= \,n\,|n \rangle $.

In other words, in the infinitely-dimensional space of fields
there is one direction
parametrized by the variable
${\cal K}=\int K_0 d^3 x$ which forms a closed circle.
The wave function (as a function of ${\cal K}$) is the
Bloch superposition
(\ref{tvac}). The parameter $\theta$ is nothing but a ``quasimomentum"
\cite {Gribov,Jackiw,DGross}.
In QCD it is called the vacuum angle.

In this formulation the $\t$ angle enters as an arbitrary phase
in (\ref {tvac}) and varies in the interval from 0 to $2\pi$.
Physics must be $2\pi$ periodic in $\theta$.
At $\theta\neq 0$ or $\pi$ one can expect $P$ and $T$
noninvariant effects.

It is straightforward to
show that for any {\it gauge invariant} operator
${\hat {\cal O}}$
\beq
\langle \t^{\prime} |\, {\hat {\cal O}}  \, |\t \rangle \,\sim \,
\delta_{\t^{\prime}\,\t}\,.
\label{tvac1}
\eeq
Therefore, no gauge invariant operator can transform
a state of one $\t$ world into a state of another $\t$ world.
The different $\t$ worlds are disconnected
from each other.

\subsection{Constraints on {\boldmath $\t$}}

As was mentioned, nonvanishing
  $\t$  leads to
$CP$ violating observables in QCD. (The
point $\t =\pi$ will be discussed
separately).  It is known that strong interactions conserve
$CP$.   Hence, a natural
question arises as to  what are the experimental constraints
on the value of $\t$.

In a full theory, with quarks, there is an additional
contribution to the
CP odd part of the Lagrangian. It  comes from the imaginary phases
of the quark mass matrix ${\cal M}$. These phases can be rotated
away from the mass matrix by chiral transformations of the quark fields.
However, because of the axial anomaly \cite {Anomaly},
which manifests itself as a  noninvariance of the
Feynman integral measure under the chiral transformations
at the quantum level  \cite {Fujikawa},
the quark mass matrix phases appear in front of the
$G\tilde G$ term in the QCD Lagrangian. Therefore,
the actual parameter that sets  the magnitude of
$CP$ violation in QCD is
\beq
\t\,+\,{\rm arg}\left ( {\rm det}\,{\cal M}  \right )\,.
\label{thetashift}
\eeq
In what follows, for  simplicity of notation,
we will denote this parameter by the same letter $\t$,
implying that   the  part
${\rm arg}\left ( {\rm det}\,{\cal M}  \right )$ is included by default.

Perhaps, the most pronounced effect of the $\t$ term is
 generation of a nonzero neutron electric dipole moment (nEDM).
The latter is parametrized by
the following  effective Lagrangian
\beq
{\cal L}_{nEDM}\,=\,{d_n^{\gamma}\over 2}\,{\bar n}\,i\,\gamma_5
\,\sigma_{\mu\nu}\,n\,F^{\mu\nu}\,,
\label{ nedm}
\eeq
where the photon field strength is
$F_{\mu\nu}\,=\,\partial_\mu\,{\cal A}_\nu \,-\,\partial_\nu\,{\cal A}_\mu$,
and $n$ stands for the  neutron. Moreover,  $\sigma_{\mu\nu} $ is the
antisymmetric product of two Dirac's gamma matrices,
$\sigma_{\mu\nu}\,\equiv\,[\gamma_\mu\,\gamma_\nu]/2i$.

In the presence of the $\t$ term,
  nEDM can be  found from the following matrix element:
\beq
\langle \,n(p_f)\,\gamma(k)\,|e\,J^{\rm em}_\mu\,{\cal A}^\mu\,\cdot \,
i\,\int\,d^4x\, \Delta \,{\cal L}_{\t}\,|n(p_i)\,\rangle\,=\,
{d_n^{\gamma}}\,{\bar n}(p_f)\,\gamma_5
\,\sigma_{\mu\nu}\,n(p_i)\,k^\mu \epsilon ^\nu(k) \,,
\label{nEDM1}
\eeq
where   $ J^{\rm em}_\mu $ is the quark electromagnetic
current. The momentum carried by the photon,
$k_\mu =(p_f)_\mu - (p_i)_\mu$, equals to the difference between
the final and initial momenta of the neutron while
$\epsilon ^\nu(k)$ denotes the photon polarization four-vector.

The matrix element on the left-hand side of (\ref {nEDM1})
is a  highly nonperturbative object. Its calculation
in QCD is nontrivial.
Nevertheless, there are a number of different methods
by which   nEDM had been estimated in the past. We will list them below.
The bag model calculation was performed in Ref.
\cite {Baluni}. The result is
\beq
d_n^{\gamma}|_{\rm Bag}\,\simeq\,\t\,\,2.7\cdot 10^{-16}\,e\cdot {\rm cm}\,.
\label{Baluni}
\eeq
Shortly after Ref. \cite {Baluni}, the chiral
logarithms (CL) method was used
 \cite {Crewther} leading to the estimate
\beq
d_n^{\gamma}|_{\rm CL}
\,\simeq\,\t\,\,5.2 \cdot 10^{-16}\,e\cdot {\rm cm}\,.
\label{Crewther}
\eeq
The method of chiral perturbation theory (ChPT)
was further advanced in Ref. \cite {Pich} with
the following result
\beq
d_n^{\gamma}|_{\rm ChPT}
\,\simeq\,\t\,\,3.3\cdot 10^{-16}\,e\cdot {\rm cm}\,.
\label{Pich}
\eeq
Finally, the most recent paper on QCD sum rule (SR) calculations \cite {SVZSR}
of $d_n^{\gamma}$ gives \cite{Maxim}
\beq
d_n^{\gamma}|_{\rm SR}\,\simeq\,
\t\,\,1.2\cdot 10^{-16}\,e\cdot {\rm cm}\,.
\label{Maxim}
\eeq
All  results above have a considerable uncertainty, of  at least $50\%$,
which  reflects a variety of uncertainties
inherent to  nonperturbative QCD calculations.
Even though the results scatter by a factor of several units, it
is beyond any doubt that $d_n^{\gamma}|_{\rm theor}\sim \t \, 10^{-16}\,  e\cdot{\rm cm}$.

This  number  should be compared with
the most recent experimental result for nEDM
presented in Ref. \cite {Harris}
\beq
|d_n^{\gamma}|_{\rm exp}\,<\,6.3 \cdot 10^{-26}\,e\cdot {\rm cm}\,.
\label{Exp}
\eeq
One gets a very strong
constraint on the value of $\t$,
\beq
|\t|\lsim  10^{-9}\,.
\label{texp}
\eeq

We see that if the value of $\t$ is nonvanishing it has to be
unnaturally small. There is no {\it a priori} reason why
two terms in (\ref {thetashift}),
the bare theta parameter and
${\rm arg}\left ( {\rm det}\,{\cal M}  \right )$,
should cancel each other
with such an extraordinary accuracy, of one part in $10^{9}$ or better.
A dynamical mechanism is needed to
explain  the  unnatural smallness of the $\theta$  term.

Before proceeding further, let us mention that
  other $CP$ odd effects are induced by the $\t$ term too.
They impose less stringent bounds on $\t$, however.
For instance, a nonzero $\t$ gives rise to a nonvanishing amplitude of
a $CP$ violating decay $\eta\,\to \pi^+\,\pi^- $ for which
${\rm Br}(\eta\,\to \pi^+\,\pi^-)\,\simeq\,\t^2\,\,1.8\cdot 10^2$
\cite {SVZ,Pich}.  The experimental limit for this decay is
${\rm Br}_{\rm exp}(\eta\,\to \pi^+\,\pi^-)\,<\,1.5\cdot 10^{-3}$.
This yields  a constraint $|\t|\,<\,3\cdot 10^{-3}$,
 much weaker than (\ref {texp}).

\subsection{Can QCD solve the strong ${\rm CP}$ problem itself ?}

Thus, theorists' task is to try to find a  mechanism
which would
make $CP$ conservation in strong interactions
natural. Two alternative approaches are logically possible.
One can  invoke physics beyond QCD (this approach
  will be discussed in the
bulk  of this review) or one can try
 a minimalistic standpoint and
ask whether QCD  could solve
the strong $CP$ problem itself,  with no new physics.

An obvious solution of the latter type
 exists:  were  one of the quarks massless, e.g., $m_u=0$, then all
$\t$ effects would be unobservable.
In this case there is a global $U_A(1)$ symmetry of the
chiral rotations of the $u$ quark field,
$u_R\to {\rm exp}(i\beta)\,u_R\,,~\,u_L\to {\rm exp}(-i\beta)\,u_L\,,$
which can eliminate $\t$ altogether.
However, $m_u=0$ does not go through on  phenomenological grounds
\cite {mu}, and at present  this scenario may be safely discarded.

A more intricate solution could exist if confinement itself were
to ensure    the effects of the $\t$ term to be screened.
As far as we know, this question was first raised by A. Polyakov
shortly after the discovery of the strong $CP$ problem.
His argumentation was as follows.
The $\t$ term in the action is the integral over the full derivative,
which can be operative only if there are long-range components of the gauge fields.
In the quasiclassical approximation such components are certainly present, as is evident from instanton calculations. However, this approximation misses the most salient feature of QCD,
color confinement, which might eliminate long-range interactions (i.e. ``screen" color)
and make the integral over the full derivative vanish.
That's exactly what happens in the  (1+2)-dimensional Polyakov model of
color confinement \cite{Polyakov:1976fu}.

 This issue was studied in
Ref. \cite {SVZ}, with the  negative conclusion.
The problem is that there are two effects in QCD which depend on the
above ``screening": the heaviness of the $\eta '$ mass (the so-called
U(1) problem) and $CP$ conservation/violation.
Even though we do not know precisely
how exactly the confinement mechanism works in QCD, we know for a fact
that $\eta '$ is split from the octet of the Goldstone bosons.
This knowledge (plus some reasonable arguments
regarding the value of chiral and/or $1/N_c$ corrections) is sufficient to show
that confinement in QCD does not eliminate
the $\t$ dependence and thus does not solve the strong $CP$ problem.
We briefly outline
this line of reasoning  below.

The flavor singlet meson, the  $\eta'$, is significantly
 heavier than the flavor octet Goldtones,
$m_{\eta'}\approx 958\,{\rm MeV}$. As was shown by Weinberg in the pre-QCD era,
were the $\eta'$ a Goldstone, its mass would be constrained by
$m_{\eta'}< \sqrt 3 m_\pi$. This suggests that,
unlike the octet of the Goldstone bosons, the  $\eta'$
is not massless in the chiral limit
(unless chiral expansion is invalid).
This is the only fact we will need.

An extra contribution to the
$\eta'$ mass comes from nonperturbative
effects due to the axial anomaly, as was exemplified
\cite {tHooft} by the same instantons.

To quantify the effect on the theoretical side,
let us introduce the correlator of
the topological charge densities (in the literature it is
referred to as
the topological susceptibility)
\beq
{\cal X}\,=\,-\,i\,\int\,d^4x\,\langle 0| T\,Q(x)\,Q(0)\,|0 \rangle \,,
\label{chi}
\eeq
where
\beq
Q\,\equiv\,{1\over 32\,\pi^2}\,G_{\mu\nu}^a\,
{\tilde G}_{\mu\nu}^a\,.
\label{dens}
\eeq

According to the low-energy theorem derived in Refs. \cite {Witten,Veneziano}
in the leading approximation of the $1/N_c$ expansion, the quantity
${\cal X}$ is saturated by the $\eta'$ contribution implying
the following formula for
the $\eta'$ mass:
\beq
m_{\eta'}^2\,=\,{6\,{\cal X} \over f_{\pi}^2}\,+\,{\cal O}\left(m_q \right )\,
+\,{\cal O}\left({1\over N_c^2}\right )\,,
\label{VW}
\eeq
where the topological susceptibility ${\cal X}$ on the right-hand side
is evaluated in pure gluodynamics, the Yang-Mills theory with
 no light quarks.
In order for the $\eta'$ mass to be nonzero in the chiral limit,
${\cal X}$ should be nonzero in pure gluodynamics.

A substantial amount of theoretical evidence is accumulated
in the last 20 years
showing
that the topological susceptibility in pure Yang-Mills   does not vanish.
The lattice \cite {Latticechi} and the QCD sum rule
studies \cite {SRchi} yield
${\cal X}\,\simeq \,(180~{\rm MeV})^4\,\ne\, 0$.
This successfully takes care of the $U(1)$ problem.

Having nonzero topological susceptibility in
pure gluodynamics means, {\em per se},  that there is a sensitivity
to the parameter $\t$ in this theory.
Indeed, ${\cal X}$ is nothing but the
second derivative of the vacuum energy with respect to $\t$
taken at $\t=0$
\beq
{\cal X}\,=\,-\,{\partial^2 \over \partial\,\t^2} \,
\left ( \left.  { \ln \, Z_{\t}\over V}
\right )\right|_{\t=0}\,.
\label{Z}
\eeq

In the theory with the light quarks   included, the topological susceptibility
can be calculated by applying the chiral perturbation theory (see e.g. \cite {SVZ}).
As expected on general grounds, in this case
${\cal X}\propto  m_q$ provided the $\eta '$ is split from the octet of the
Goldstones; otherwise the $\eta '$ contribution cancels
the $O(m_q)$ term in the topological susceptibility.
However, in  the theory with the light quarks it is much more instructive
to calculate directly $CP$ odd decay rates, for instance
the rate of $\eta\,\to \pi^+\,\pi^- $. This amplitude is forbidden by $CP$.
In the same way as with the topological susceptibility,
the chiral low-energy theorem yields \cite {SVZ} a
nonvanishing amplitude $O(m_q)$
provided the $\eta '$ is split from the octet of the
Goldstones.

Therefore, since QCD does solve the  U(1)  problem --- it does
split the $\eta '$ from the octet Goldstones ---  it cannot solve the
strong CP problem \cite{SVZ} without help from outside \footnote{
The situation seems to be quite clear in this respect,
nevertheless, an attempts to develop
models of confinement that would solve
the strong CP problem  are not abandoned.
Let us  comment on a proposal of Ref.
\cite {Samuel} where two distinct ``topological susceptibilities"
are defined: the local and global ones.
Let $V_c$ denote the volume at which the confinement effects take place,
and   $V$ be the total volume of space time $V_c \ll V\to \infty$.
According to \cite {Samuel}, the local topological susceptibility is
$
{\cal X}_{\rm loc}\,=\,
\int_{V_c}\,d^4x\,\langle 0| T\,Q(x)\,Q(0)\,|0\rangle \,,
$
while the global  one  is
$
{\cal X}_{\rm glob}\,=\,\int_V\,d^4x\,\langle 0| T\,Q(x)\,Q(0)\,
|0\rangle \,.
$
The author claims that the  solution of the  U(1)  problem
requires that ${\cal X}_{\rm loc}\ne 0$,
while the  solution of  the strong $CP$ implies ${\cal X}_{\rm glob}= 0$,
and both  conditions may be  dynamically compatible, so that
both the
strong $CP$ and U(1) problems   could be solved simultaneously.
The underlying dynamics outlined in \cite {Samuel} is a
  special interaction  between
instantons   which ``screens" them.

There are a number of objections to this suggestions.
First,  it is the global topological susceptibility
that enters in
the Witten-Veneziano relation
and   determines the $\eta'$ mass modulo $1/N_c$ corrections.
Even if we forget about theoretical calculations of this quantity
demonstrating that it does not vanish,
we know that the $\eta '$ is split from the
Goldstone octet because the Weinberg relation
$m_{\eta'}< \sqrt 3 m_\pi$ is grossly violated.
(Of course, the validity of the chiral expansion is assumed;
otherwise the Weinberg relation is meaningless.)
If the $\eta '$ is split, there is no way out \cite{SVZ}:
$\t$-induced effects are observable.

In terms of the model suggested in \cite {Samuel}
this means that if one were able to complete the
calculations at the level of physical observables,
one would find that the $\eta '$ is not split from
the octet of the  Goldstones
or, more likely, that the required interaction between
instantons is not sustainable.
The latter variant was advocated in \cite {Dowrick}. }.

\section{In search of  a solution beyond QCD}

\subsection{Peccei-Quinn mechanism}

The first dynamical mechanism   solving the strong $CP$
problem was proposed by Peccei and Quinn \cite {PQ}.
The main observation of Ref. \cite {PQ} is as follows:   if there is
a  U(1)  axial symmetry in the theory
\beq
q_{L}\, \to \, e^{i\alpha}\,q_{L}\,,~~~~~q_{R}\, \to \,
e^{-i\alpha}\,q_{R}\,,
\label {PQ}
\eeq
then the $\theta$  term can be removed from the Lagrangian,
much like in  the case of one massless quark
discussed in the previous section.
Below  this symmetry will be referred to as
U(1)$_{\rm PQ}$.

To see whether this symmetry is present in the
Standard Model with one Higgs doublet
let us consider the Yukawa sector
and restrict ourselves to the first generation  quarks
(consideration in the  general case is quite similar),
\beq
\lambda_u\,{\bar Q}_{L}\,\phi\,u_{R}\,+\,\lambda_d\,
{\bar Q}_{L}\,\phi_c\,d_{R}\,+\,{\rm H.c.}\,+\,V(\phi^+\,\phi )\,,
\label {SMY}
\eeq
where $Q_L$ is the left-handed SU(2)$_W$ quark
doublet, $(\phi_c)_i\,=\,\epsilon_{ij}\,\phi^{*j}$ is
the charge conjugate Higgs field, and $V$ denotes the Higgs potential.
Although the first term in this expression is invariant under
the transformations
\beq
q_{L}\, \to \, e^{i\alpha}\,q_{L}\,,~~~q_{R}\, \to \, e^{-i\alpha}\,q_{R}\,,
~~~\phi\,\to \,e^{2i\alpha}\,\phi\,,
\label {PQPhi}
\eeq
there is a second term which is not invariant
under (\ref {PQPhi}) since $\phi_c$ transforms as conjugate to
$\phi$.  Therefore, the one Higgs doublet SM   is {\em not} invariant under
U(1)$_{\rm PQ}$ and the strong $CP$ cannot be solved.

However, as was pointed out in Ref. \cite {PQ},
the required U(1)$_{\rm PQ}$   symmetry is present in  SM
with {\it two} Higgs doublets ---  let us call them
$\phi$ and $\chi $. In this case the Yukawa sector
for the first generation quarks reads
\beq
\lambda_u\,{\bar Q}_{L}\,\phi\,u_{R}\,+\,\lambda_d\,
{\bar Q}_{L}\,\chi^*\,d_{R}\,+\,{\rm H.c.}\,+\,V(\phi^+\,\phi\,,~
\chi^+\chi\,,~(\phi^+\,\chi)\,(\chi^+\phi))\,.
\label {SMPQ}
\eeq
It is invariant under
\beq
q_{L}\, \to \, e^{i\alpha}\,q_{L}\,,~~~q_{R}\, \to \, e^{-i\alpha}\,q_{R}\,,
~~~\phi\,\to \,e^{2i\alpha}\,\phi\,,~~~\chi\,\to \,e^{-2i\alpha}\,\chi\,.
\label {PQPhichi}
\eeq
This fact can be used to solve the strong $CP$ problem \cite {PQ}.
The symmetry (\ref {PQPhichi}) is explicitly broken by
the axial anomaly. As a result,
instantons induce an effective potential
for the $\theta$  term. The potential can be calculated in a
certain approximation \cite {PQ}. The crucial model-independent
fact is that the resulting potential is minimized  at a zero value of the
$CP$ violating  phase \cite {PQ}.

\subsection{Weinberg-Wilczek axion}

When the Higgs fields develop vacuum expectation values
(VEV's) the local electroweak symmetry group is spontaneously
broken. This gives  masses to
the intermediate $W^{\pm}$ and $Z$ vector bosons.
Simultaneously, the global $U(1)_{\rm PQ}$ is spontaneously
broken too. Spontaneous breaking of the {\it global}
symmetry leads to the emergence of a massless Goldstone
boson, the axion in the present case \cite {Weinberg, Wilczek}.
In   SM with two Higgs doublets the axion
is given by the following superposition
\beq
a\,\equiv\,{1\over v}\,\left (v_{\phi} {\rm Im}\phi_0\,-\,
v_{\chi} {\rm Im}\chi_0\,\right )\,,
\label {PQaxion}
\eeq
where $\phi_0$ and $\chi_0$ denote the neutral
components of the Higgs doublets. Moreover,
$v\,\equiv\, \sqrt{v_{\phi}^2\,+\,v_{\chi}^2}\simeq 250\,{\rm GeV}$,
and $v_{\phi}  $ and   $v_{\chi}  $ are the vacuum expectation
values of $\phi$ and $\chi$, respectively.
In this approximation  the axion is massless. However, as we mentioned above,
  nonperturbative QCD effects (such as instantons) give rise
to a potential for the axion. Hence, the axion acquires a nonzero mass
which can be estimated as follows \cite {Weinberg,Wilczek}
\beq
m_a\,\simeq \,{f_\pi\,m_\pi \over v}\, \simeq \, 100\,{\rm KeV}\,.
\label {maWW}
\eeq
Moreover, the axion decay constant is $1/v$.
Therefore, we see that the Weinberg-Wilczek (WW) axion mass and
decay constant are tied to the electroweak
symmetry breaking scale $v$.  This turns out to be too much of a constraint,
and, as we will discuss in Sec. 3.5, the
WW axion is excluded on the basis of existing experimental data.

\subsection{KSVZ axion}

If the scale of PQ symmetry breaking is
much higher than the electroweak scale, then
according to (\ref {maWW}),
the axion is  much lighter and its
decay constant is  much
smaller.  Such an ``invisible'' axion would
not be in  conflict with   experimental data.

A scenario with the harmless axion was first proposed
in Refs. \cite {Kim} and \cite {SVZ} (the KSVZ axion).
In the latter paper it was called {\em phantom axion}.
Needless to say that to untie the axion from the electroweak scale
one has to decouple the corresponding scalar fields from the
known quarks and couple them to hypothetical (very) heavy
fermion fields
carrying color.

In more detail, one
 introduces a complex scalar field
$\Phi$ coupled to a hypothetical electroweak singlet, a quark field $Q$ in the
fundamental representation of   color SU(3),
\beq
\Delta {\cal L} = \Phi \bar Q_R Q_L + \mbox{H.c.}\,.
\eeq
The modulus of $\Phi$ is assumed to develop
a large vacuum expectation value $f/\sqrt{2}$, while the argument of
$\Phi$
becomes the axion field $a$, modulo normalization,
\beq
a(x) = f \alpha (x) \,,\quad \alpha (x)\equiv \mbox{Arg} \Phi (x)\,,
\qquad f\gg\gg \Lambda\,.
\eeq
Then the low-energy  coupling of the axion to the gluon field is
\beq
\Delta {\cal L} = \frac{1}{f} \, a\, \frac{1}{32\pi^2}\,
G_{\mu\nu}^a\tilde G_{\mu\nu}^a\, ,
\label{axglc}
\eeq
so that the QCD Lagrangian depends on the
combination $\theta + \alpha (x)$.

In general, one could  introduce more than one
fundamental field $Q$, or introduce them  in a higher representation
of the color group. Then, the axion-gluon coupling (\ref{axglc})
acquires an integer multiplier $N$,
\beq
\Delta {\cal L}' = \frac{1}{f} \, a\, N\, \frac{1}{32\pi^2}\,
G_{\mu\nu}^a\tilde G_{\mu\nu}^a\,.
\label{axglcp}
\eeq
This factor $N$ (not to be confused with
the number of colors $N_c$ nor with ${\cal N}$ of extended supersymmetry)  is sometimes referred to as the axion index. The minimal
axion corresponds
to $N=1$. In the general case the QCD Lagrangian  depends on the
combination $\theta + N\alpha (x)$. As previously,
  nonperturbative  QCD effects generate a potential
for $\theta + N\alpha (x)$. The latter is minimizes
at the value $\theta + N\alpha_{\rm vac}\, =\,0$, i.e.,
the strong $CP$ problem is automatically solved.

\subsection{ZDFS axion}

An alternative way to introduce an ``invisible''
axion was proposed in Refs. \cite {Zhitnitsky} and
\cite {DFS}, (the ZDFS axion).
In this proposal one maintains the PQ symmetry
of the two doublet SM but separates the scales of
the PQ and electroweak breaking \cite {Zhitnitsky,DFS}.
To   this end the SM Lagrangian is extended --
a scalar SM singlet field $\Sigma$ is added,
\beq
&& \lambda_u\,{\bar Q}_{L}\,\phi\,u_{R}\,+\,\lambda_d\,
{\bar Q}_{L}\,\chi^*\,d_{R}\,+\,{\rm H.c.}\,+\nonumber \\[3mm]
&& V(\phi^+\,\phi\,,~
\chi^+\chi\,,~(\phi^+\,\chi)\,(\chi^+\phi)\,,~\Sigma^+\Sigma\,,~
(\phi^+\chi)\,\Sigma^2)\,.
\label {ZDFS}
\eeq
One notes that this expression is invariant under the following
axial transformations
\beq
q_{L}\, \to \, e^{i\alpha}\,q_{L}\,,~~q_{R}\, \to \, e^{-i\alpha}\,q_{R}\,,
~~\phi\,\to \,e^{2i\alpha}\,\phi\,,~~\chi\,\to \,e^{-2i\alpha}\,\chi\,,~~
\Sigma\,\to \,e^{2i\alpha}\,\Sigma \,.
\label {PQZDFS}
\eeq
Upon spontaneous breaking of this symmetry the
Goldstone particle, an axion,
emerges as a   superposition
\beq
a\,\equiv\,{1\over V}\,\left (v_{\phi} {\rm Im}\phi_0\,-\,
v_{\chi} {\rm Im}\chi_0\,+\,  v_{\Sigma}\, {\rm Im}\Sigma
\right )\,,
\label {ZDFSaxion}
\eeq
where $V\,\equiv\, \sqrt{v_{\phi}^2\,+\,v_{\chi}^2\,+\,
v_{\Sigma}^2 }$,
and $v_{\phi}  $, $v_{\chi}  $  and $v_{\Sigma}$
are the vacuum expectation values of $\phi$, $\chi$ and
$\Sigma$, respectively. The vacuum expectation value of
$\Sigma$ does not have to be related to
the electroweak symmetry breaking scale. In fact,
it can be as large  as the GUT scale. If so, the
axion is light and its decay constant is
tiny. We will discuss experimental bounds on these quantities in
the next section.

\subsection{Constraints on the axion mass}

As we discussed in the previous sections,
the PQ symmetry is explicitly broken by the axial anomaly.
Therefore, the axion is a {\it pseudo} Goldstone boson.
Nonperturbative QCD effects induce the axion mass.
For further discussions it is convenient to parametrize the axion
mass as follows
\beq
m_a\,\simeq\,0.6\,{\rm eV}\,{10^7\,{\rm GeV}\over f}\,,
\label{axionmass}
\eeq
where $f$ is the axion decay constant
determined by the PQ breaking scale.

In general, while discussing phenomenological constrains on
the axion mass, one should distinguish between the KSVZ and ZDFS
cases.  The axion couplings to matter
are different in these two scenarios.
In particular, the KSVZ axion has no tree-level couplings to
the standard model quarks and leptons.
However, the aim of the present
section is to summarize briefly an order of magnitude
constraints on the axion mass. For this goal
the effects which distinguish between the KSVZ and ZDFS axions
will  not be important (for detailed studies
see Ref. \cite {Raffelt} and citations therein).

The quantity $1/f$ sets the strength of the axion coupling.
Light axions can  be produced in stars and
a part of the energy of a star can be carried
away by those  axions. Stars can loose energy due to the production
of light axions in the following possible processes

(i) Nucleon-nucleon bremsstrahlung:
$N\,+\,N\,\to \, N\,+\,N\,+\,a\,;$

(ii) The Primakoff process:
$\gamma \,\leftrightarrow\,a$ conversion in
the electromagnetic field of a nucleus;

(iii) Photoproduction on an electron:
$\gamma \,+\,e^{-}\,\to\,e^{-}\,+\,{a}\,;$

(iv) Electron bremsstrahlung on a nucleus:
$e^-\,+\,(A,Z)\,\to \,e^-\,+\,a\,+\,(A,Z)\,;$

(v) Photon fusion: $\gamma\,+\,\gamma \,\to\,a\,;$

Unless  $1/f$ is really small, the
emission of axions that are produced in the above reactions
would lead to unacceptable energy loss by the star.
This leads to the following
lower bound on the axion decay constant
$f\, \gsim \, 10^9\,{\rm GeV}$.

It is remarkable that cosmology puts an
{\it upper} bound on $f$ \cite{C1,C2,C3}.
The latter comes about as follows.
If $f$ is too large then
the axion coupling $1/f$ is very small.
As a result, during the course of cosmological evolution
of the universe axions decouple early and
begin to oscillate coherently.
There are two major mechanisms by which the
energy density stored in these oscillations can be
dissipated -- the Hubble expansion of the universe and
the particle production by axions.
However, if $f\,\gsim \, 10^{12}\,{\rm GeV}$,
neither of  these mechanisms are effective (the
axion coupling is  too small).
As a result, at some point of the evolution
the axion energy density exceeds the critical energy density
and over-closes the universe. In order for this not to happen
one  should impose the constraint $f \,\lsim \, 10^{12}\,{\rm GeV}$.

Summarizing, we obtain the following order of magnitude
bounds  on $f$ and $m_a$:
\beq
10^9\,{\rm GeV}\,\lsim \, f \,\lsim \,10^{12}\,{\rm GeV}\,,~~~
10^{-6}\,{\rm eV} \,\lsim\,m_a \,\lsim \, 10^{-3}\,{\rm eV}\,.
\label{fama}
\eeq
For further details see, e.g., Ref. \cite {Raffelt}.

\section{The vacuum structure in large $N_c$ gluodynamics}

The early studies \cite{Crewther:1977ce} of the chiral Ward
identities in QCD revealed that the vacuum energy density
depends on the vacuum angle $\theta$ through the ratio
$\theta /N_f$, where $N_f$ is the number of   quarks with mass
$m_q\ll\Lambda$.  Shortly after it was shown in Refs.
\cite{Witten:1980sp} and \cite{DiVecchia:1980ve} that this
structure occurs  naturally, provided that  there exist  $N_f$
states in the theory such that one of them  is the true vacuum,
while others are  local extrema; all are intertwined in the
process of ``the $\theta$ evolution."  Namely, in passage from $\theta
=0$ to  $\theta = 2\pi$, from   $\theta = 2\pi$ to  $\theta = 4\pi$, and
so on, the roles of the above states interchange: one of the local extrema
becomes the global minimum  and {\em vice  versa}. This would imply,
with necessity, that at $\theta = k\pi$ (where $k$ is an odd integer)
there are two degenerate vacuum states. Such a group of  intertwined
states will be referred to as the ``vacuum family." The crossover at
$\theta = \pi$, $3\pi$, etc. is called the Dashen phenomenon
\cite{DASH}.

This picture was confirmed by a detailed examination of
effective  chiral Lagrangians
\cite{Witten:1980sp,DiVecchia:1980ve,Rosenzweig,Arnowitt}  (for a
recent update see \cite{Smilga:1999dh}).  For two and three light
quarks with equal masses  it was found that the vacuum family
consists of two or three states respectively; one of them is a global
minimum  of the  potential, while others are local extrema.\footnote{We
stress that the states from the vacuum family need not necessarily lie
at the  minima of the energy functional. As was shown by Smilga
\cite{Smilga:1999dh}, at certain values of $\theta$ some may be
maxima. Those which intersect at $\theta = k\pi$ ($k$ odd) are
certainly the minima at least   in the vicinity of $\theta =k\pi$.}  At
$\theta =\pi$ the levels intersect.  Thus, Crewther's dependence
\cite{Crewther:1977ce}  on  $\theta /N_f$  emerges.

On the other hand, the examination of the effective chiral Lagrangian
with the realistic values of the quark masses,
$m_d/m_u \sim 1.8\,,\,\,\, m_s/ m_d \sim 20$, yields
\cite{Witten:1980sp,DiVecchia:1980ve,Smilga:1999dh} a drastically
different picture --  the vacuum family disappears (shrinks to one
state); the crossover phenomenon at $\theta=\pi$ is gone as well.

This issue remained in a dormant state for some time. Recently
arguments were given  that the ``quasivacua" (i.e. local minima of the
energy functional), which together with the true vacuum form a
vacuum family,  is an indispensable feature of gluodynamics. The first
argument in favor of this picture  derives \cite{Shifman:1997ua}
from supersymmetric gluodynamics, with supersymmetry softly broken
by a gluino mass term. The same conclusion was reached in Ref.
\cite{Witten:1998uk} based on  a D-brane construction in the limit of
large $N_c$. One can see that in both approaches
the number of states  in the vacuum family scales as $N_c$.
In fact, in Ref.  \cite {Witten:1998uk} the expression
for the theta dependence of the vacuum energy  in the large $N_c$ pure
Yang-Mills (YM) model was derived from a D-brane construction.
It has the following form \cite
{Witten:1998uk}\footnote {It has been conjectured long time ago in \cite
{Witten:1980sp}.}:
\begin{eqnarray}
{\cal E}(\theta) =C ~
{\rm min}_k~(\theta + 2\pi k)^2~+ {\cal O} \left
( {1\over N_c}
\right ),
\label{energy}
\end{eqnarray}
where $C$ is some  constant independent of $N_c$ and $k$ stands for
an integer number. This expression has a number of
interesting features which might seem a bit puzzling from the
field theory
point of view. Indeed, in the large $N_c$ limit there are
$N_c^2$ degrees of
freedom in  gluodynamics, thus, naively, one would expect that the vacuum
energy density  in this theory scales as $\sim N_c^2$.
However, the leading term in Eq. (\ref {energy}) scales as $\sim 1$.
As a natural explanation, one could conjecture that there should be
a colorless massless excitation which saturates the expression for the
vacuum energy density (\ref {energy}).
However, pure gluodynamics generates a mass gap and there
are no  physical massless excitations in the model. Thus, the
origin of Eq. (\ref {energy}) seems to be a conundrum.
We will discuss and resolve  this puzzle in the next section.

Note, that an additional argument in favor of the
vacuum family may be found in a cusp
structure which develops once one sums up \cite{Halperin:1998bs}
sub-leading in $1/N_c$ terms in the effective $\eta '$ Lagrangian.
At large $N_c =\infty $ the states from the vacuum family are stable,
and so are the domain walls interpolating between them
\cite{Witten:1998uk,ShifmanM}.

When $N_c <\infty$ the degeneracy and the vacuum  stability
is gone, strictly speaking. It is natural to ask what happens
if one switches on the axion field.
This generically leads to the formation of the axion domain walls.
The axion domain wall
\cite{Sikivie:1982qv} presents  an excellent set-up for studying
the properties of the QCD vacuum under the $\theta$ evolution.
Indeed, inside the axion wall, the axion field (which, in fact, coincides
with an effective $\theta$) changes  slowly from zero to $2\pi$.
The characteristic
length scale, determined by the inverse axion mass
$m_a^{-1}$, is huge in the units  of QCD, $\Lambda^{-1}$.
Therefore, by  visualizing a  set of spatial slices parallel to the axion
wall,
separated by distances $\gg \Lambda^{-1}$, one obtains
a  chain of  QCD laboratories with  distinct values of $\theta_{\rm
eff}$ slowly varying from one slice to another. In the middle of the wall
$\theta_{\rm eff} =\pi $.

Intuitively,
it seems clear that in the middle of the axion wall,
the effective value of $\theta_{\rm eff} =\pi$.
Thus, in the central part of the wall
 the hadronic sector is effectively in the regime with
two degenerate vacua, which entails a {\em stable} gluonic wall
as a core of the axion wall. In fact,
we deal here with an axion wall ``sandwich."
Its core is the so-called D-wall, see \cite{GabadShifman}.

Below we will discuss this idea more thoroughly.
We also address the question  whether this phenomenon
persists  in the theory with  light quarks, i.e.,
in real QCD. Certainly, in the limit $N_c = \infty$
the presence of quarks is unimportant, and the axion wall
will continue to contain the D-wall core.
As we lower the number of colors, however,
below  some critical number
it is inevitable that the regime must change,
the gluonic core must disappear as a result of  the absence of the
crossover.  The parameter governing the change of the regimes
is $\Lambda /N_c$ as compared to the quark mass $m_q$.
At $m_q \ll \Lambda /N_c$, even if one forces the
axion field to form a wall, effectively it is screened by a dynamical
phase whose origin can be traced to the $\eta '$, so that in the
central part of the axion wall the hadronic sector does not develop
two degenerate vacua. The D-walls cannot be accessed in this case
via the axion wall.

The issue of hadronic components of the axion wall in
the context of a potential with cusps \cite {Halperin:1998bs}
were discussed in \cite
{Fugleberg:1999kk,Halperin:1998gx,Forbes:2000gr}.
However, the gluonic component of the axion walls  was  not studied.
The $\eta'$ component in the axion walls was  discussed in \cite
{Evans:1997eq,Fugleberg:1999kk}.

\subsection{Arguments from supersymmetric gluodynamics}

First
 we will summarize arguments in favor of
the existence of a nontrivial vacuum family
in pure gluodynamics.

The first indication
that the crossover phenomenon
may exist in gluodynamics
comes \cite{Shifman:1997ua}
from supersymmetric Yang-Mills theory,
with supersymmetry being broken by a gluino mass term.
The same conclusion was reached in Ref. \cite{Witten:1998uk}
based on  a D-brane construction in the limit of large $N_c$.
In both approaches
the number of states  in the vacuum family is $N_c$.

The Lagrangian of
softly broken supersymmetric gluodynamics is
\beq
L &=& \frac{1}{g^2}\left\{ -\frac{1}{4} G_{\mu\nu}^a G_{\mu\nu}^a
+ i\, \bar\lambda^a_{\dot\alpha}
D^{\dot\alpha\alpha}\lambda^a_{\alpha}
-\left( m \lambda^a_{\alpha}\lambda^{a\alpha}+\mbox{H.c.}\right)
\right\}
\nonumber \\[0.2cm]
& +&
\theta \, \frac{1}{32\pi^2}\, G_{\mu\nu}^a
\tilde G_{\mu\nu}^a\,,
\label{alsgl}
\eeq
where $m$ is the gluino mass which is assumed to be small, $m\ll
\Lambda$ (here we rescaled the gluon and gluino  fields
so that $1/g^2$  appears as a common multiplier in the Lagrangian).

There are $N_c$ distinct chirally asymmetric vacua,
which (in the $m=0$ limit) are labeled by
\beq
\langle \lambda^2\rangle_\ell = N_c\Lambda^3 \exp\left(i \,
\frac{\theta +2\pi
\ell}{N_c}
\right)\,,\quad \ell = 0,1,..., N_c-1\,.
\eeq
At  $m=0$ there are stable  domain walls interpolating between them
\cite{DvaliShifman}.
Setting $m\neq 0$  we eliminate the
vacuum degeneracy.
To  first order in $m$ the vacuum energy density
in this theory is
\beq
{\cal E} = \frac{m}{g^2}\langle \lambda^2\rangle + \mbox{H.c.}
= - m\,N_c^2 \Lambda^3 \, \cos \frac{\theta +2\pi
\ell}{N_c}\,.
\eeq
Degeneracy of the vacua is gone.
As a result, all the metastable vacua
will decay very quickly.
Domain walls between them, will be
moving toward infinity because of the finite
energy gradient between two adjacent vacua.
Eventually one ends up with a single true vacuum state in the
whole space.

For each given value of $\theta$ the ground state energy is given by
\beq
{\cal E} (\theta ) = {\rm min}_{\ell}
\left\{ - m\,N_c^2\,\Lambda^3 \, \cos \frac{\theta +2\pi
\ell}{N_c}\right\}\,.
\label{dfour}
\eeq
At $\theta = \pi$, $3\pi$, ..., we observe the vacuum degeneracy and
the crossover phenomenon. If there is no phase transition in $m$,
this structure will survive, qualitatively,
even at large $m$ when the gluinos disappear from the spectrum,
and we recover pure gluodynamics.

Based on a D-brane construction Witten showed \cite{Witten:1998uk}
that in pure SU($N_c$) (non-supersymmetric) gluodynamics
in the limit $N_c\to\infty$
a vacuum family does exist:\footnote{This was shown in Ref.
\cite{Witten:1998uk} assuming that there is no phase transition
in a certain parameter of the corresponding D-brane construction.
In terms of gauge theory, this assumption amounts of saying that
there is no phase transition as one interpolates
to the strong coupling constant regime.
Thus, the arguments of \cite{Witten:1998uk}
have the same disadvantage as those of SUSY gluodynamics
where one had  to assume the absence of the phase transition in
the gluino mass.} the theory has  an infinite group  of states
(one is the true vacuum,
others are non-degenerate metastable ``vacua'')
which are intertwined  as $\t$ changes by $2\pi\times ({\rm
integer})$, with a crossover at $\t = \pi\times$(odd integer).
The energy density of the $k$-th state from the family is
\beq
{\cal E}_k(\theta) ~= N_c^2 ~\Lambda^4~
F\left(\frac{\theta + 2\pi k}{N_c}\right)\,,
\label{energ}
\eeq
where $F$ is some $2\pi$-periodic function, and
the truly stable vacuum for each $\theta$ is obtained  by minimizing
${\cal E}_k$ with respect to $k$,
\beq
{\cal E}(\theta) ~= N_c^2\,  \Lambda^4\,\mbox{min}_k\,
F\left(\frac{\theta + 2\pi
k}{N_c}\right)\,,
\label{energg}
\eeq
much in the same way as in Eq. (\ref{dfour}).

At very
large $N_c$ Eq. (\ref{energg}) takes the form
\begin{eqnarray}
{\cal E}(\theta)~ =~\Lambda^4 ~{\rm min}_k~(\theta + 2\pi k)^2~+ {\cal
O} \left
( {1\over N_c}
\right )~.
\label{energyN}
\end{eqnarray}
The energy density ${\cal E}(\theta)$ has its
absolute minimum at $\theta =0$. At $N_c =\infty$ the ``vacua"
belonging to  the vacuum family
are stable but non-degenerate.
To see that the lifetime  of the metastable ``vacuum"
goes to infinity  in
the large $N_c$ limit one can consider
the domain walls which separate these vacua
 \cite {ShifmanM,GabadC}. These walls are  seen as wrapped
D-branes in the
construction of \cite {Witten:1998uk}, and they indeed resemble many
properties of the QCD D-branes on which  a QCD string could end.
We refer to them as D-walls because of  their striking similarity to
D2-branes.  The  consideration of D-walls
has been carried out  \cite {ShifmanM} and leads to the conclusion
that  the lifetimes of the quasivacua  go to infinity
as ${\rm exp} (\mbox{const}\, N_c ^4)$.

Moreover, it was argued \cite {DGK,GabadShifman} that the
width of these wall  scales as $1/N_c$ both, in SUSY and pure
gluodynamics.
To reconcile this observation  with
the fact that masses of the  glueball mesons scale as $N_c^0$,
we argued  \cite {GabadShifman} that there should exist
heavy (glue)
states with  masses $\propto N_c $ out of  which the walls are built.
The  D-brane analysis \cite{PStr}, effective Lagrangian arguments
and  analysis of the wall junctions \cite{Gorsky:2000hk},
support this interpretation. These heavy states resemble properties
of D0-branes. The analogy is striking, as D0-branes make
D2-branes from the standpoint of the  M(atrix) theory \cite {BFSS},
so these QCD ``zero-branes'' make QCD D2-branes (domain
walls).\footnote{See also  closely related discussions in Ref.  \cite
{GabadKakush}.}
The distinct vacua from the vacuum family
differ from each other by a restructuring of these  heavy
degrees of freedom. They
are essentially  decoupled from the glueballs in the large $N_c$ limit.

Now we switch on the axion
\beq
\Delta{\cal L} = \frac{1}{2}f^2 (\partial_\mu \alpha )
(\partial^\mu \alpha ) +\frac{\alpha }{32\pi^2}\, G_{\mu\nu}^a
\tilde G_{\mu\nu}^a\,,
\eeq
with the purpose of studying the axion walls. The potential energy
${\cal E} (\theta)$ in Eq. (\ref{energg}) or (\ref{energyN}) is replaced by
${\cal E} (\theta +\alpha)$.

Since the hadronic sector exhibits a nontrivial
vacuum family
and the crossover\footnote{For nonminimal axions,
with $N\geq 2$, the crossover occurs at  $\alpha = k\pi /N$.}
at $\theta = \pi, 3\pi$,
etc., strictly speaking, it is impossible to integrate out completely the
hadronic
degrees of freedom in studying the axion walls.
If we want to resolve the cusp, near the cusp we have to deal with the
axion field
{\em plus} those hadronic degrees of freedom which restructure.
In the middle of the wall, at  $\alpha = \pi$,
it is mandatory to jump from one hadronic vacuum to another
-- only then the energy of the overall field configuration
will be minimized and the wall be stable.
Thus, in  gluodynamics the axion wall acquires a D-wall
core by necessity.

One can still
integrate out the heavy   degrees of freedom
everywhere except a narrow strip (of
a hadronic size) near the middle of the wall.
Assume for simplicity that
there are two states in the hadronic family.  Then the low-energy
effective Lagrangian for the axion field takes the form (\ref {axYM}).
The domain wall profile will also exhibit
a cusp in the second  derivative. The wall solution takes the form:
\beq
\alpha (z ) =
\left\{\begin{array}{l}
8 \, {\rm arctan}\, \left( e^{m_a z}\,{\rm tan}\frac{\pi}{8}
\right)\,,\quad \mbox{at $z< 0$}\\[0.3cm]
-2\pi +8 \, {\rm arctan}\, \left( e^{m_a z}\,{\rm tan}\frac{3\pi}{8}
\right)\,,\quad \mbox{at $z> 0$}\,,
\end{array}
\right.
\label{dwc}
\eeq
at $N_c= 2$ (the wall center is at $z=0$).

Examining this cusp with
an appropriately high resolution
 one would observe that it is smoothed on the
hadronic  scale, where the hadronic component of the
axion wall ``sandwich'' would
become visible.
The cusp carries a finite contribution to the
wall tension which cannot be calculated in the
low-energy approximation but can be readily estimated,
$T_{\rm core} \sim \Lambda^3N_c$.
In subsequent sections
we will  examine this core manifestly in a toy solvable model.

\subsection{Argumens from D-brane construction}

The theta dependent vacuum energy (\ref {energy})
is related to the correlator measuring the vacuum fluctuations of
the topological charge. The question which arises here is whether this can be
seen from the original string theory computation \cite
{Witten:1998uk}. We are going to discuss below  how the
string theory calculation suggests that the vacuum energy
(\ref {energy}) should indeed be related to the vacuum fluctuations of the
topological charge. In fact, we argue that this is related to the instantons
carrying $D0$-brane charge in the Type IIA fourbrane construction of the
four-dimensional YM model.

In general, a great deal of information can be learned
on nonperturbative phenomena in four-dimensional gauge theories
by obtaining these models as a low-energy
realization of certain D-brane configurations \cite {Wittenbranes}, and/or
using  a duality of large $N$ superconformal gauge
theories and string theory compactified on certain spaces (see Refs. \cite
{Maldacena} and \cite {GKPolyakov,WittenAdS}).
This duality, being a powerful technique,
has also been generalized for the case of non-supersymmetric models \cite
{Witten1}. This was applied to study various dynamical issues in large $N_c$
pure Yang-Mills theory \cite
{Gross,Ooguri1,Ooguri2,Klebanov,KlebanovTseytlin,Minahan}.

To begin with let us recall how the theta dependent vacuum energy  appears in
the brane construction of the four-dimensional YM model \cite {Witten:1998uk}.
One starts with Type IIA superstring theory on
${\cal M}\equiv {\bf R}^4\times {\bf S}^1\times {\bf R}^5$,
with $N_c$ coincident $D4$-branes \cite {Witten1}.
The $D4$-brane worldvolume is assumed to be $ {\bf R}^4 \times {\bf S}^1$
and the fermion boundary conditions on ${\bf S}^1$
are chosen in such a way that the low-energy theory on the worldvolume is
pure non-supersymmetric $U(N_c)$ YM theory \cite {Witten1}.
In the dual description,
the large $N_c$ limit of the $SU(N_c)$ part of this theory can be studied by
string  theory on a certain background \cite
{Maldacena,GKPolyakov,WittenAdS,Witten1}. It was shown in Ref. \cite
{Witten:1998uk} that the theta dependent vacuum energy (\ref {energy})
arises in the dual string description due to  the  $U(1)$ gauge field
$B_M,~M=1,..,5$. To find out what this corresponds to in the original gauge
theory language  recall that this $U(1)$ field is nothing but the
Ramond-Ramond (RR) one-form of Type IIA theory. Furthermore, once the gauge
theory is realized in the  Type IIA fourbrane construction, the
Wess-Zumino-Witten (WZW) term present in the worldvolume effective action
defines the correspondence between the gauge theory operators on one
side and the string theory Ramond-Ramond fields on the other side.
In the case at
hand the worldvolume WZW term  looks as follows:
\begin{eqnarray}
{S}_{\rm WZW}={1\over 8\pi^2}~\int_{\Omega}~B\wedge {\rm Tr}\,G\wedge G,
\label{WZW}
\end{eqnarray}
where ${\Omega}$ denotes the worldvolume of a  wrapped fourbrane,
${\Omega}\equiv {\bf R}^4\times {\bf S}^1$. In accordance with
the general principles of the large $N_c$ AdS/CFT correspondence \cite
{Maldacena,GKPolyakov,WittenAdS} the
classical action for the RR one-form  on the string theory
side defines the YM correlation
functions of the composite operator $G{\tilde G}$
(since this is the operator which couples to the corresponding RR field in
(\ref {WZW})). Thus, it is not
surprising that the theta dependent vacuum energy which is defined by the RR
one-form in the string theory calculation is related to the nonzero value of
the topological susceptibility in the gauge theory studies.
The physical reason for this correspondence, as we have mentioned above,
is the special property of the gauge theory instantons in
the fourbrane construction.
Indeed, in accordance with (\ref {WZW}) the RR one-form
couples to the topological charge density $G{\tilde G}$, on the other hand
the RR one-form couples by definition to $D0$-branes. Thus, the gauge theory
instantons in this case carry zerobrane charge. This is  the physical reason
for the correspondence discussed above.

\subsection{Derivation of vacuum energy in QCD}

The aim of this section is to derive
Eq. (\ref {energy}) in pure YM model and, in particular,
to identify the degrees of freedom which are responsible for the theta
dependent vacuum energy density.

In the quasi-classical  approach the theta
dependence can be  calculated using instantons \cite {Polyakov}. In a
simplest approximation  of non-interacting instantons the theta
angle enters the Euclidean space partition function in the following form:
\begin{eqnarray}
\exp \Big (- {8\pi^2\over g^2} \pm i\theta  \Big )\equiv \exp\Big ( -N_c
{8\pi^2\over \lambda} \pm i\theta \Big )~,
\label{theta}
\end{eqnarray}
where $g$ stands for the strong coupling
constant. $\lambda$ denotes the 't Hooft's coupling $\lambda\equiv
N_c~g^2$ which
is kept fixed in the large $N_c$ limit.  The expression above vanishes in the
large $N_c$ limit, so does the theta dependence in (\ref {theta}). However, this
conclusion cannot be extrapolated to the infrared region
of the model. The limitations of the expression (\ref
{theta}) prevent one to do so. Indeed, the quasi-classical approximation is
valid in the limit of small coupling constant (see, for instance, discussions
in Ref. \cite {Coleman}). Once
quantum corrections are taken into account the coupling constant  $g^2$ in
(\ref {theta}) becomes a  scale dependent quantity. In fact, it will depend
upon an instanton size $g^2=g^2(\rho)$. For small size instantons the
running coupling is small and the quasi-classical approximation in (\ref
{theta}) holds. However, for large size instantons, i.e. large
couplings, it is not even clear whether the notion of a single instanton is a
legitimate approximation. The overlap  between instantons can be big in this
case and some more complicated field configurations  should be
relevant for the description of physical phenomena \cite {Shuryak}.
In any event, the expression (\ref {theta})
is no longer reliable in  the strong coupling limit.
Thus, the conclusion that the theta dependence goes away in the large $N_c$ limit
cannot be justified. One way to study the infrared region is to look for some
appropriate composite colorless excitations for which the notion of  an
asymptotic state can be used. We will start by searching for these
excitations in pure Yang-Mills theory. To proceed, let us recall
that the topological susceptibility, ${\cal X}$, is a nonzero
number in pure gluodynamics (we rewrite it in the following form)
\begin{eqnarray}
{\cal X} = -i~\int \partial^\mu ~\partial^\nu \langle
0|T~K_{\mu}(x)K_{\nu}(0)|0\rangle~d^4x\neq 0~.  \label{Chi}
\end{eqnarray}
Here, $K_\mu $ as before
denotes the Chern-Simons current. As we discussed earlier,
the value of ${\cal X}$ in large $N_c$ pure
YM theory determines the $\eta'$ meson mass in full QCD with massless quarks
via the  Witten-Veneziano formula
$m^2_{\eta'} f^2_{\eta'}\propto {\cal X}$, with $f_{\eta'}$ being the $\eta'$
meson decay constant \cite {Witten,Veneziano}\footnote {
Below, unless otherwise stated,
we will not distinguish between $\cal X $ and its large $N_c$ limit.
The constant contact term in the definition of $\cal X$
will also be omitted for simplicity.}.

In what follows it will prove convenient   to  introduce  a new variable
by rewriting  the expression for the topological charge density
$Q$ in terms of a four-index (four-form) tensor field
$H^{\mu\nu\alpha\beta}$:
\begin{eqnarray}
Q = { \varepsilon_{\mu\nu\alpha\beta}H^{\mu\nu\alpha\beta}
 \over 4!}~, \label{QH}
\end{eqnarray}
where the four-form field $H^{\mu\nu\alpha\beta}$ is the field
strength for the three-form potential $C_{\mu\nu\alpha}$:
\begin{eqnarray}
H_{\mu\nu\alpha\beta}=\partial_\mu C_{\nu\alpha\beta}-
\partial_\nu C_{\mu\alpha\beta}-\partial_\alpha C_{\nu\mu\beta}-
\partial_\beta C_{\nu\alpha\mu}.  \label{HC}
\end{eqnarray}
The $C_{\mu\nu\alpha}$ field is defined as a
composite operator of the gluon fields $A^a_\mu$:
\begin{eqnarray}
C_{\mu\nu\alpha}={1\over 16 \pi^2}(A^a_\mu
{\overline {\partial}}_\nu
A^a_\alpha-A^a_\nu {\overline {\partial}}_\mu A^a_\alpha-A^a_\alpha
{\overline {\partial}}_\nu
A^a_\mu+ 2 f_{abc}A^a_\mu A^b_\nu  A^c_\alpha). \label{CA}
\end{eqnarray}
Here, $f_{abc}$ denote the structure constants of the corresponding $SU(N_c)$
gauge group.  The right-left  derivative in this expression is defined as
$A{\overline {\partial}}B\equiv A (\partial B)-(\partial A) B $.
Notice, that the $C_{\nu\alpha\beta}$ field is not a gauge invariant
quantity; if the gauge transformation parameter is
$\Lambda^a$, the three-form field transforms as
$C_{\nu\alpha\beta}\rightarrow
C_{\nu\alpha\beta}+\partial_\nu \Lambda_{\alpha\beta}-
\partial_\alpha \Lambda_{\nu\beta}-\partial_\beta
\Lambda_{\alpha\nu}$,
where $\Lambda_{\alpha\beta}\propto A_\alpha ^a\partial_\beta
\Lambda^a - A_\beta  ^a\partial_\alpha \Lambda^a$. However, one can
check that the expression for the field strength $H_{\mu\nu\alpha\beta}$
is gauge invariant.

It has been known for some time
\cite {Luscher} that the $C_{\nu\alpha\beta}$  field
propagates  long-range correlations if
the topological susceptibility is nonzero in the theory.
The easiest way to see this is to turn to
the notion  of the Kogut-Susskind pole \cite {KogutSusskind}.
Let us consider the correlator of the vacuum topological susceptibility at a
nonzero momentum.
In this case ${\cal X} $ is defined
as the zero momentum limit of the correlator of two Chern-Simons
currents multiplied by two momenta:
\begin{eqnarray}
{\cal X} =-i~\lim_{q\rightarrow 0}q^\mu q^\nu \int e^{iqx}\langle 0|T
K_\mu (x) K_\nu (0)|0\rangle d^4x.
\label{chinew}
\end{eqnarray}
Since this expression is nonzero, it must be that the correlator of two
Chern-Simons  currents develops a pole
as the momentum vanishes, the Kogut-Susskind pole \cite
{KogutSusskind}.

Given that the correlator of two Chern-Simons  currents
has a pole  and that
the Chern-Simons current and the three-form
$C_{\nu\alpha\beta}$  field  are related,
one concludes  that the $C_{\nu\alpha\beta}$ field also has a nonzero
Coulomb  propagator \cite {Luscher}.
Thus, the $C_{\nu\alpha\beta}$ field behaves as a
massless collective excitation propagating
a long-range interaction  \cite {Luscher}.
These  properties, in the large $N_c$ limit,
can  be  summarized in the following
effective action  for
the $C_{\nu\alpha\beta}$ field:
\begin{eqnarray}
S_{eff}=-{1\over 2\cdot 4!~{\cal X} }\int
H^2_{\mu\nu\alpha\beta}~d^4x
-{\theta \over 3!} \int_{\partial \Gamma} C_{\nu\alpha\beta}
~dx^\nu \wedge dx^\alpha \wedge dx^\beta +
{\rm High~dim.}
\label{action}
\end{eqnarray}
The first term in this expression
yields  the correct Coulomb propagator
for the three-form $C_{\nu\alpha\beta}$ field. The second term is just the
usual ${\rm CP}$  odd  $\theta$ term of the initial YM action
written as a surface integral at spatial infinity $\partial \Gamma$.
Notice that higher dimensional  terms are not explicitly
written in this expression. There might be two types of higher dimensional
contributions in
(\ref {action}). First of all, there are terms with higher powers of
derivatives of the fields. These terms are suppressed by momenta of the
``massless'' three-form field and do not contribute to the zero momentum vacuum
energy of the system.  In addition, there might be higher dimensional terms
with no additional derivatives. In the next section we will present some of
these contributions and show that they are suppressed by higher powers of
$1/N_c$.

In what follows we are going to study the
large $N_c$ effective action  given in Eq. (\ref
{action}) \footnote {The  action (\ref {action})
is not an  effective action in the Wilsonian sense. It is rather
related to the generating functional of one-particle-irreducible
diagrams of the composite field in the large $N_c$ limit. The
effective action in Eq. (\ref {action})  is not to be quantized  and loop
diagrams of that action  are not to be taken into
account in calculating  higher order Green's functions.
The analogous effective action for
the $\rm CP$ even part of the theory was constructed in
Refs. \cite {Schechter},  \cite{MigdalShifman}.}.
In particular,  we will calculate
the ground state  energy of the system in the large $N_c$ limit
using the effective action (\ref
{action}). In fact, we will derive Eq. (\ref {energy}).

Before we turn to this
calculation let us mention that Maxwell's equations for a  free four-form
field-strength  $H_{\mu\nu\alpha\beta}$ yield only a constant solution in
$(3+1)$-dimensional
space-time \cite {Aurilia}. The reason  is as follows.
A four-form potential has only one independent degree of freedom
in four-dimensional space-time, let us call it $\Sigma$. Then,
the four Maxwell's equations written in terms of the $\Sigma$ field
require  that this field is independent of the all four space-time
coordinates, thus the solution can only be a space-time constant.
As a result, the free $H_{\mu\nu\alpha\beta}$ field propagates
no dynamical degrees of freedom in $(3+1)$-dimensions.
However, this field can be
responsible for a positive vacuum energy density
in various  models of Quantum Field Theory (see Ref. \cite {Townsend}).
Thus, studying classical equations of motion for the
$H_{\mu\nu\alpha\beta}$ field one can  determine
the value of the ground state energy given by configurations of
$H_{\mu\nu\alpha\beta}$.
We are going to solve explicitly the classical equations  of motion
for the effective action (\ref {action}). Then,
the energy density associated with those solutions will be calculated.

Let us start with the equations of motion. Taking the variation of the
action (\ref {action}) with respect to the $C_{\nu\alpha\beta}(z)$
field one gets
\begin{eqnarray}
\partial^\mu H_{\mu\nu\alpha\beta}(z)=\theta~ {\cal X}~
\int_{\partial \Gamma}  \delta^{(4)}(z-x)
~ dx_\nu \wedge dx_\alpha \wedge dx_\beta .
\label{equationmotion}
\end{eqnarray}
This
equation  can  be solved exactly in four-dimensional space-time \cite
{Aurilia}. The solution is the sum of a particular solution of the
inhomogeneous equation and a  general solution of the
corresponding  homogeneous equation:
\begin{eqnarray}
H_{\mu\nu\alpha\beta}(z)=\theta ~{\cal X}~
\int \delta^{(4)}(z-x)
~dx_\mu \wedge  dx_\nu \wedge dx_\alpha \wedge dx_\beta +b~
\varepsilon_{\mu\nu\alpha\beta}.
\label{constant}
\end{eqnarray}
The integration constant $b$, if nonzero,  induces an  additional
CP violation beyond the existed  $\theta$ angle.
However, periodicity of the $\theta$ angle
with respect to shifts  by $2\pi \times ({\rm integer})$ allows
for some nonzero $b$ proportional to  $2\pi {\bf Z}$.
As a result, the general solution to  the equation of motion
reads as follows:
\begin{eqnarray}
H_{\mu\nu\alpha\beta}=-(\theta +2\pi k)~{\cal X}~
\varepsilon_{\mu\nu\alpha\beta}~.
\label{Htheta}
\end{eqnarray}
Thus, the different vacua are labeled by the integer $k$ and the order
parameter for these vacua in the large $N_c$ limit
can be written as:
\begin{eqnarray}
\langle G{\tilde G} \rangle _k=~(\theta +2\pi k)~{\cal X}.
\label{order}
\end{eqnarray}
As a next step let us
compute the vacuum energy associated with
the solution given in Eq. (\ref {Htheta}). The density of the
energy-momentum tensor for the action (\ref {action})
takes the form
\begin{eqnarray}
\Theta_{\mu\nu}=-{1\over 3!~ {\cal X}    }\left (
H_{\mu\alpha\beta\tau}H_{\nu}^{~\alpha\beta\tau}-{1\over 8}
g_{\mu\nu} H^2_{\rho\alpha\beta\tau}\right ). \label{emt}
\end{eqnarray}
Using the expression (\ref {Htheta})
one calculates the corresponding
energy density\footnote {
Notice that the total YM energy density should contain some  negative
constant related to the nonzero value of the gluon condensate
\cite {SVZSR}.
This constant is subtracted from the expression for the energy discussed in
this work. The energy density (\ref {newenergy}) is normalized as
${\cal E}_0(\theta=0)=0$, and
for $k=0$ was discussed in \cite {membrane}.}  ${\cal E}_k$
\begin{eqnarray}
{\cal E}_k(\theta)={1\over 2}(\theta+2\pi k)^2~ {\cal X}.
\label{newenergy}
\end{eqnarray}
Since the $H_{\mu\nu\alpha\beta}$ field does not propagate
dynamical degrees of freedom the expression above is the
total energy density of the system given by the action  (\ref {action})
\footnote {One  might wonder
whether the same result is obtained if one treats
$\theta$ not as a constant multiplying  $Q$ in the Lagrangian,
but as the phase that the states acquire under a topologically
non-trivial gauge transformations.
In this case the arbitrary integration constant
in Eq. (\ref {constant}) has to be chosen in such a way
which would guarantee a
proper $\theta$ dependence of the VEV of the topological charge density.
This would leave the results of our discussion without modifications.}.

Before we go further let us stop here to discuss some of the consequences
of Eq. (\ref {newenergy}). First of all, let us notice that
the result (\ref {newenergy}), as well as Eq. (\ref {energy}),
is only valid  in the
limit of infinite $N_c$. Below we  will calculate
subleading order corrections to  Eqs. (\ref {energy},\ref {newenergy})
and argue that these expressions can also  be used as a good approximation
for large but finite $N_c$. The constant $C$ emerging
in (\ref {energy}) is related to the topological susceptibility as follows:
$$ {\cal X}|_{N_c\rightarrow \infty}=2 C.$$
Thus, the vacuum energy (\ref {energy},\ref {newenergy})
is defined by vacuum fluctuations of
the topological charge measured by $\cal X$.

The crucial feature of (\ref {newenergy}) is that it defines an infinite number
of vacua. The true vacuum is obtained by minimizing  (\ref
{newenergy}) with respect to $k$ as in (\ref {energy}):
\begin{eqnarray}
{\cal E}_0(\theta)={1\over 2}~ {\cal X}~{\rm min}_k~
(\theta+2\pi k)^2~.
\nonumber
\end{eqnarray}
This expression is periodic with respect to shifts of $\theta$ by
$2\pi {\bf Z}$ and is also a smooth function of $\theta$ except
for $\theta=\pi$ \cite {Witten:1998uk} (see also discussions below).
Thus, there are an
infinite number of the false vacua in the theory \cite {Witten:1998uk}.
The fate of these states will be discussed in section 5.

Let us now consider full QCD with three quark flavors. We are going to
write down a low-energy effective Lagrangian for this case and then gradually
decouple quarks by taking the quark masses to infinity. The resulting effective
Lagrangian should be giving the energy density for pure Yang-Mills theory.

In the  large $N_c$ expansion the effective Lagrangian of
QCD with three flavors takes the form \cite
{VenezianoDi,Rosenzweig,Arnowitt}: \begin{eqnarray}
{\cal L}(U,U^*,Q)={\cal L}_0(U,U^*)+
{1\over 2 }~ i~ Q (x)~ {\rm Tr}~ \Big ( {\rm ln}~U-{\rm ln}~U^*\Big )+
\nonumber  \\
{1\over 2~{\cal X}}~ Q^2(x) +\theta~ Q(x) +{B \over 2 \sqrt{2} }~{\rm Tr}
~(MU+M^*U^*)+\dots ~,
\label{effective}
\end{eqnarray}
where $U$  denotes the flavor group matrix of pseudoscalar mesons,
${\cal L}_0$ denotes the part of the Lagrangian which
contains the meson fields only \cite
{VenezianoDi,Rosenzweig,Arnowitt},
$B$ is some  constant related to the quark condensate,
and $M$ stands for the meson mass matrix (for recent discussions of
the effective chiral Lagrangian approach see Ref. \cite {PichRev}).
Higher order
terms in (\ref {effective}) are suppressed by quadratic and higher powers of
$1/N_c$.  In order to study vacuum properties, we concentrate
on the low-momentum approximation.
The Lagrangian presented above can be used to solve the $U(1)$ problem \cite
{Witten,Veneziano}. Indeed, the field $Q$ enters the Lagrangian in a quadratic
approximation and can be integrated out. As a result, the flavor singlet
meson, the $\eta'$, gets an additional contribution into its mass term.
This leads to the Witten-Veneziano relation and the solution of the $U(1)$
problem without instantons \cite {Witten,Veneziano}. In the present case we
would like to follow an opposite way. Namely, we are going to make quarks
very heavy and integrate them out keeping the field $Q$ in the Lagrangian.
In the limit $m_q\rightarrow \infty$ one finds that $M\rightarrow
\infty$. Thus, the low-energy effective Lagrangian which is left after the
mesons are integrated out will take the form:
\begin{eqnarray}
{\cal L}_{\rm eff} (Q)=
{ 1\over 2~{\cal X} }~ Q^2(x) +\theta~ Q(x) +
{\cal O}~ \Big ( { \Lambda^2_{\rm QCD} \over M^2 },~~{\partial^2 Q^2\over
\Lambda^{10}_{\rm QCD}},~~{1\over N_c^2} \Big )~.  \label{effectiveYM}
\end{eqnarray}
Rewriting the field $Q$ in terms of the ``massless''
tensor $C_{\alpha\beta\gamma}$ as in
the previous section,  one finds that
the expression (\ref {effectiveYM}) is nothing but the Lagrangian presented in
(\ref {action}). Thus, the higher order terms neglected in
(\ref {action}) which could contribute to the vacuum energy at zero
momenta would correspond to higher corrections in $1/N_c$. In fact, the
subleading  corrections to  the effective Lagrangian (\ref {effective}) can
also be found \cite {Pettorino}. These terms are proportional (with the
corresponding dimensionful coefficients) to the following expressions:
\begin{eqnarray}
{ {\rm const.} \over N_c^2 }~ Q^2~{\rm Tr}~
(\partial_\mu U~\partial_\mu U^*),
~~~~~~~~~~~{ {\rm const.}\over N_c^2 }~ Q^4~.
\label{corrections}
\end{eqnarray}
The terms listed above are suppressed in the effective Lagrangian
by the factor $1/N_c^2$.
As a next step, we can include the terms (\ref {corrections}) into the full
effective Lagrangian and then integrate the heavy meson fields out. The net
result of this procedure is that the terms proportional to $Q^4$ appear in the
effective Lagrangian for pure YM theory. This, in its turn,
modifies the equation of motion for the single component
of $H_{\mu\nu\alpha\beta}$ considered in the previous section.
 Performing the calculation of the vacuum
energy in the same manner as discussed above we find the following result
for the energy density:
\begin{eqnarray}
{\cal E}_k(\theta)={1\over 2}~ {\cal X} ~\Big (
\theta+2\pi k\Big )^2+ {{\rm const.}\over N_c^2}~ {\cal X}~
\Big (  \theta+2\pi k\Big )^4~+~{\cal O} \Big ( {1\over N_c^3} \Big ).
\label{1overN}
\end{eqnarray}
In this expression a constant emerges as a result of integration
of the equation  of motion\footnote {The
numerical value of this constant was recently calculated on the lattice
\cite {newlattice}.}.
Notice that the topological susceptibility in the
expression above is also defined in the corresponding order in the large
$N_c$ expansion: ${\cal X}=2 C+{\cal X}_{1}/N_c+{\cal X}_2/N_c^2$.
Thus, the expressions (\ref {newenergy},\ref {1overN}) could in
principle  give a reasonable approximation for big enough but otherwise {\it
finite} $N_c$.
The true vacuum energy density, ${\cal E}_0(\theta)$, can be obtained by
minimizing the  expression (\ref {1overN}) with respect to $k$ as in (\ref
{energy}). Then, ${\cal E}_0(\theta)$ satisfies the relation
$\partial^2_{\theta}
{\cal E}_0(\theta)~|_{\theta=0} ={\cal X} $, no matter what
is the value of the constant in (\ref {1overN}).

\section{Dynamics of false vacua}

In this section we will discuss the dynamics of the false vacua present
in the theory. In accordance  with (\ref {newenergy},\ref {1overN})            there are an infinite number of vacua for any given value of
the theta angle.  Clearly, not all of these are degenerate.
As we discussed,
the true vacuum state is defined by minimizing the expressions
(\ref {newenergy},\ref {1overN})
with respect to $k$. All the other states are false
vacua with greater values of the energy density. There is a potential barrier
that separates a given false vacuum state from the true one. Thus, a false
vacuum can in general  decay into the true state through the process of
bubble nucleation \cite {bubble}\footnote {This decay can go through the
Euclidean ``bounce'' solution \cite {bounce}. Though the  existence of
the bounce for this case is not easy to understand  within the field theory
context, nevertheless, one could  be motivated by the brane construction where
this object appears as a sixbrane bubble wrapped on a certain
space \cite {Witten:1998uk}.}. The decay rates for these vacua  were
evaluated in  Ref. \cite {ShifmanM}.  In this section
we analyze  the fate of the false vacua
for different realizations of the initial conditions in which the system is
placed. For the sake of simplicity  we will be discussing transitions between
the vacuum states labeled by $k'$ and $k$ for different  values
of these integers.
The first two cases considered in this section were studied in Refs. \cite
{Witten:1998uk} and \cite {ShifmanM}, the remaining of the section follows
Ref. \cite {GabadC}.

\subsection {False vacua with $k'\sim 1$}

In this subsection we consider the system which in its initial state
exists in a
false vacuum with $k'$ of order  $\sim 1$. Let us start with the case
when $N_c$ is a large but finite number so that the formula (\ref
{newenergy}) (or (\ref {1overN})) is still a good approximation. Since there
exists the true vacuum state with less energy, the false vacuum can ``decay''
into  the true one
via the bubble nucleation process. That is to say, there is a finite
probability to form
a bubble with the true vacuum state inside.  The shell of the bubble
is a  domain wall which separates the false state from the true one.
The dynamical question we discuss here is whether it is
favorable energetically to create and expand such a bubble.
Let us study the energy balance for the case at hand. While creating the shell
of the bubble one looses the amount of energy equal to  the surface area
of the bubble multiplied by the tension of the shell. On the other hand, the
true vacuum state is created inside the bubble, thus, one gains  the amount of
energy equal to the difference between the energies of the false and true
states.  The energy balance between these two effects defines whether
the bubble can be formed, and, whether the whole false vacuum
can transform into
the true one by expanding  this bubble to infinity.
Let us start with the volume energy density.
The amount of the  energy density which is gained
by creating the bubble is\footnote{In this subsection we
assume that $\theta
\neq \pm \pi$. The case  $\theta=\pi$ will be considered below.}:
\begin{eqnarray}
\Delta {\cal E} ={1\over 2}~ {\cal X}~
\Big [ (\theta +2 \pi)^2-\theta^2 \Big ]=2
\pi~ {\cal X}~ (\theta+\pi).
\nonumber
\end{eqnarray}
Thus, $\Delta {\cal E}$    scales as $\sim 1$ in the large $N_c$ limit as
long as the volume of the bubble is finite. Let us now
turn to the surface energy which is lost. This energy
is defined as:
\begin{eqnarray}
{E}_s= T_D~\times ({\rm surface ~~area})~.
\label{surface}
\end{eqnarray}
The tension of the wall between the adjacent vacua, $T_D$, should scale as
$T_D\sim N_c$ in the large $N_c$ limit. Hence, the surface
energy will also scale as $\sim N_c$. Thus, the process of creation of
a finite volume bubble in the  large
$N_c$ limit is not energetically favorable.
Indeed,
the amount of energy which is lost while creating the shell
is bigger than the amount which is gained.
In terms of the false vacuum decay width
this means that the width of this process is suppressed in the
large $N_c$ limit
\cite {ShifmanM}:
\begin{eqnarray}
{\Gamma\over {\rm Volume} }~\propto ~ \exp \Big (- a N_c^4 \Big ),
\label{width}
\end{eqnarray}
where $a$ stands for some positive constant \cite {ShifmanM}.
Thus, one concludes that in the limit $N_c\rightarrow\infty$  the false vacua
with $k'\sim 1$ are stable \cite {Witten:1998uk,ShifmanM}.

\subsection {False vacua with $k'\sim N_c$}

Here we study the fate of the false vacua with $k'\sim N_c$. We discuss
a possibility that these vacua  can decay into a state $k$
with $k'-k\sim N_c$ and $k'+k\sim N_c$.  As in the previous subsection, we are
going to study the energy balance for the bubble nucleation process. The amount
of the volume energy density which is gained by creating  such a bubble
in the large $N_c$ limit scales as follows:
\begin{eqnarray}
\Delta {\cal E}~\propto ~ {\cal X}~ N_c^2.
\nonumber
\end{eqnarray}
Thus, the volume energy which is gained increases as $\sim N_c^2$. Let us now
turn to the surface energy which is lost while nucleating a bubble. This is
defined as ${E}'_s=T'_D\times ({\rm surface~~area})$, where $T'_D$
denotes the  tension of the domain wall interpolating between the vacua
labeled by $k'$ and $k$. Since $k'-k\sim N_c$ these vacua are not neighboring
ones. Thus, in general, there is no reason to expect that the tension of the
wall interpolating between these vacua scales as $\sim N_c$.
$T'_D$ might scale
as $\sim N_c^2$ at most (as the energy of a generic
configuration in a model  with $N_c^2$ degrees of freedom).
However, even in the case when $T'_D\sim N_c^2$
the volume energy which is gained is at least of the same order
as the  surface energy which is lost. Hence,
it is energetically favorable to increase the radius of such a bubble
(since the volume energy scales as a cubic power of the radius  while the
surface energy scales only as a quadratic power of the effective size).
Thus, the bubble nucleation process will not be suppressed and the false
vacua with $k'\sim N_c$ will eventually decay into the true ground state.
Note, that the state $k'=N_c$ can as well decay into the neighboring vacuum
$k=N_c-1$  which subsequently is allowed to turn into the ground state.

\subsection {Parallel domain walls}

In this subsection we consider the special case when all the vacua are
present simultaneously in the initial state of the model. This can be
achieved, for instance, by placing in space an infinite number of parallel
domain walls  separating  different vacua from each other. It is
rather convenient to picture these walls as parallel planes. Each vacuum state
is sandwiched between the corresponding two domain walls (two planes)
separating this state from the neighboring vacua. Each domain is
labeled by $k$ and in accordance with (\ref {newenergy},\ref {1overN}) is
characterized by the corresponding value of the vacuum energy. Furthermore,
the order parameter $\langle G{\tilde G}\rangle $ takes different values in
these vacua in accordance with (\ref {order}).  Let us turn to the true vacuum
state. For simplicity we assume  that this state is given by $k=0$ (which
corresponds to $|\theta|$  being less than $\pi$). The corresponding vacuum
energy is the lowest one.  Consider the two states which are adjacent to the
true vacuum.  These states have the energy density bigger than that of
the true vacuum.  Thus, there is a constant pressure acting on the domain walls
separating the true vacuum from the adjacent false ones. This pressure will
tend to expand the domain of the true vacuum. In fact, for large but finite
$N_c$,  the pressure will indeed expand the spatial region of the true vacuum
by  moving apart the centers of the domain walls sandwiching this
state.  The very same effect will be happening  between any two adjacent
vacua. Indeed, let us calculate the jump of the energy density between the two
vacua labeled by $k'$ and $k$:
\begin{eqnarray}
\Delta {\cal  E}_{k'k}=2\pi~ {\cal X}~ (k'-k) ~\Big (
\theta+\pi(k'+k) \Big )~.
\label{deltae}
\end{eqnarray}
As far as $N_c$ is large but finite, the walls will start to accelerate.
Farther the wall is located from the true vacuum (i.e. larger the sum
$k'+k$),  bigger the initial acceleration of the wall is going to be;
i.e., the walls will start to move apart from each other with the following
initial acceleration:
\begin{eqnarray}
a_{k'k}\propto \Lambda_{\rm  YM }~ {(k'-k)~ [\theta +\pi (k'+k)] \over N_c}~.
\label{acceleration}
\end{eqnarray}
For finite $N_c$ all the walls will be moving to spatial infinity and the
whole space will eventually be filled with the true vacuum state.
On the other hand,
when $N_c\rightarrow \infty $ the picture is a bit different.
There are a number of interesting cases to consider:

First of all let us
set $k'-k =1$ and $k'~,k \sim 1$. Then, in the limit
$N_c\rightarrow \infty $
the acceleration $a_{k'k}\rightarrow 0$.
Thus, the neighboring walls  stand still if they had no initial velocity.
The physical reason of this behavior is as follows.
Although there is a constant pressure
of order $\sim 1$ acting on the wall, nevertheless,
the wall  cannot be moved because
the mass per unit surface area of the wall
tends to infinity in the limit $N_c\rightarrow \infty$.

The second interesting case would be  when the
constant pressure produced by the energy jump between some neighboring
vacua is
of order $\sim N_c$. In this case it will  be possible to accelerate
these walls up to the speed of light and send them to spatial infinity.
Indeed, if $k'-k =1$ but $k'+k \sim N_c$, then the wall
between these two vacua starts moving with a non-vanishing acceleration
which scales as follows:
\begin{eqnarray}
a_{k'k}\propto \Lambda_{\rm YM} {\pi (k'+k) \over N_c}\sim
{\cal O} (1).  \label{nonumber}
\end{eqnarray}
Thus, these walls will eventually be approaching spatial infinity with a speed
of light even in the limit of infinite $N_c$.

In addition to  the effects emphasized above
there might also be decays of the false
vacua happening in each particular domain. As we discussed in the
previous subsections, for large but finite $N_c$ all the false vacua will be
nucleating bubbles with energetically favorable phases inside and
expanding these bubbles to infinity. Thus, for large but finite $N_c$, there are
two effects which  eliminate the false vacua: The moving walls
are sweeping these states to infinity, and, in addition, these vacua are
decaying via bubble nucleation processes.

What happens for an infinite $N_c$?
As we learned above there are an infinite number of domains which will stay
stable in that limit  and the corresponding false vacua would not decay because
of the exponential suppression. Thus, there are an infinite
number of inequivalent spatial regions which are separated
by domain walls.
Consider one of the regions sandwiched between two domain walls.
The three-form field $C_{\mu\nu\alpha}$ will couple to the walls
and the large $N_c$ effective action for this case will look as follows:
\begin{eqnarray}
{\tilde S} =S_{\rm eff}+\sum_{i=k,~k+1}~\mu_i~\int_{{\cal W}_i}
C_{\mu\nu\alpha}~dx^{\mu} \wedge dx^{\nu} \wedge dx^{\alpha},
\label{tildeS}
\end{eqnarray}
where $S_{\rm eff}$ is defined in (\ref {action}), $\mu_i$ stands
for the coupling of the three-form potential to a corresponding domain wall;
${\cal W}_i$ denotes the worldvolume of the wall. In this case the domain
wall can be regarded as a source of the corresponding three-form potential.
This is reminiscent to what happens in the large $N_c$  supersymmetric YM model
\cite {GiaZuraMe}.

\subsection {Domain walls at  $\theta=\pi$}

If $\theta=\pi$, the initial classical Lagrangian is ${\rm CP}$ invariant.
Indeed, under ${\rm CP}$ transformations $\theta=\pi$ goes into $-\pi$.
Since $\pi$ and $-\pi$ angles are equivalent,  ${\rm CP}$ is a symmetry of
the Lagrangian.  However, in accordance with (\ref {order}),
this symmetry is
spontaneously broken by the vacuum of the theory. Thus, one finds the
following two degenerate true vacua:
\begin{eqnarray}
{\cal E}_{k=0}={\cal E}_{k=-1}={1\over 2}~ {\cal X}~ \pi^2~.
\label{degenerate}
\end{eqnarray}
These two vacua are labeled by the order parameter
(\ref {order}).  In the $k=0$ state  $\langle G{\tilde G}\rangle =
\pi{\cal X}$ and in the $k=-1$ state  $\langle G{\tilde G}\rangle =
-\pi{\cal X}$.  As a result of the spontaneous breaking of a discrete symmetry
there should be a domain wall separating these two vacua.  Let us consider
the case discussed in the previous section.  Namely, let us choose the initial
condition of the system as a state where all the possible vacua are
simultaneously realized in space. That is, there are an infinite number of
domain walls (parallel planes) dividing space into an infinite number of
domains with different values of the vacuum energy density labeled
by $k$.  As we mentioned
above, only two of these domains have equal minimal energy density
given in (\ref {degenerate}).
The domain wall separating these two vacua, as we will see below, is somewhat
special. In accordance with the discussions in the
previous subsection, for large  but finite $N_c$, all the walls merging with
the false vacua will  tend to rush to spatial infinity.
The final stable state of
the model can be described as a space separated into two parts by a single
domain wall. To the left (right) of the wall one  discovers the phase with
$k=-1$ with the corresponding order parameter $\langle G{\tilde G}\rangle =
-\pi{\cal X}$ , and, to the right (left) of the wall one finds
the state with $k=0$ and $\langle G{\tilde G}\rangle =
\pi{\cal X}$.  Vacuum energies of these two states are degenerate.

In the case of infinite $N_c$ the picture is slightly different.  As elucidated
in the previous subsection, there will be an infinite number of stable vacua.
The domain wall separating  the two true vacua can be regarded in this case as
the fixed plane under ${\bf Z}_2$ transformations of the coordinate
transverse to the plane.
The  three-form field $C_{\mu\nu\alpha}$ will be able to
couple to this wall in a manner discussed in the previous subsection.

\section{Axions and vacuum structure in gluodynamics}

\subsection{Two scenarios}

The invisible axion is very light. Integrating out all other
degrees of freedom and studying the low-energy
axion effective Lagrangian must be a good approximation.
The axion effective
potential in QCD can be of two distinct types.

Assuming  that for all values of $\theta$ the QCD vacuum is unique
one  arrives at the axion effective
Lagrangian of the form
\beq
{\cal L}_a = f^2
\left[ \frac{1}{2} (\partial_\mu \alpha )^2 +m_a^2
\left(\cos (\alpha  +\theta ) - 1
\right)
\right]\, .
\label{dtwo}
\eeq
The axion potential does  not have to be
(and generically  is not) a pure cosine; it may have higher
 harmonics. In the general case it is a smooth
periodic  function of
$\alpha  +\theta $, with the period $2\pi$.
For illustration we presented the potential as a pure cosine.
This does not change the overall picture in the
qualitative aspect.

As we will see below, a smooth effective potential
of the type (\ref{dtwo})
emerges  even if  the (hadronic) vacuum family is non-trivial,
but the transition between the distinct hadronic vacua does not occur
inside the axion wall. This is the case with very light quarks, $m_q\ll
\Lambda /N_c$.  In the opposite limit, one arrives at the
axion potential with cusps, considered below.

In the theory (\ref{dtwo}) one finds
the axion walls interpolating between the vacuum state
at $\alpha = -\theta$ and the same vacuum state
at $\alpha = -\theta+2\pi$,
\beq
\alpha (z )~+~\theta = 4 \, {\rm arctan}\, ( e^{m_a z})\,,
\eeq
where the wall is assumed to lie in the $xy$ plane, so that
the wall profile depends only on $z$.
This is the most primitive ``$2\pi $ wall."

The tension of this wall is obviously  of the order of
\beq
T_1 \sim f^2 m_a\, .
\eeq
Taking into account that $ f^2 m_a^2 \sim {\cal X}$ where ${\cal X}$ is
the topological susceptibility of the QCD vacuum, we get
\beq
T_1 \sim {\cal X} /m_a\, .
\eeq
The inverse proportionality to $m_a$ is due to the
fact that the transverse size of the axion wall is very large.

Let us now discuss the axion effective  potential of the second
type.
In this case the potential has cusps, as is the case in
pure gluodynamics, where
the axion effective Lagrangian is of the form
\beq
{\cal L}_a =
 \frac{ f^2}{2} (\partial_\mu \alpha )^2
+
 \min_\ell \left\{ N_c^2 \Lambda^4
\cos \frac{\alpha +2\pi \ell}{N_c}
\right\}\, ,
\label{axYM}
\eeq
(see more detailed discussions below).
Here the $\theta$ angle was included in the definition of the
axion field. The axion wall interpolates between
$\alpha = 0$ and $\alpha = 2\pi$.

What is the origin of this cusp? The cusps reflect a restructuring in the
hadronic sector.
When one (adiabatically) interpolates in $\alpha$ from $0$ to
$2\pi$ a gluonic order parameter, for instance $\langle G\tilde
G\rangle$,  necessarily experiences  a restructuring
in the middle of the wall corresponding to the  restructuring
of heavy gluonic degrees of freedom. In other words,
one jumps from the hadronic vacuum which initially (at
$\alpha =0$) had $\langle G\tilde G\rangle = 0$ into the vacuum in
which  initially
$\langle G\tilde G\rangle \neq 0$.
Upon arrival to $\alpha =2\pi$, we find $\langle G\tilde G\rangle =
0$ again. This implies that the central part of such  an axion wall is
dominated  by a
gluonic  wall. Thus, the cusp at $\alpha = \pi$
generically   indicates the formation of
a hadronic core, the D-wall \cite{GabadShifman} in the case at hand.

Returning to the question of the tension
we
note that
\beq
{\cal X} &\sim& \Lambda^4 N_c^0\,,\quad  m_a \sim \Lambda^2
N_c^0f^{-1} \,\,\,\mbox{in pure gluodynamics}\,,\nonumber\\[0.2cm]
{\cal X} &\sim& \Lambda^3 N_c m_q\,,\quad m_a \sim
\Lambda^{3/2}m_q^{1/2} N_c^{1/2}f^{-1}
\,\,\,\mbox{in QCD with light quarks}\,,
\eeq
which implies, in turn,
\beq
T_1 \sim \left\{\begin{array}{l}
f\Lambda^2 N_c^0\,\,\,\mbox{in pure gluodynamics}\\[0.3cm]
f\Lambda^{3/2}m_q^{1/2} N_c^{1/2}\,\,\,\mbox{in QCD with light
quarks}\,.
\end{array}
\right.
\label{dthree}
\eeq
Here $m_q$ is the light quark mass.

The presence of the large parameter $f$ in $T_1$ makes the axion halo
the
dominant contributor to the wall tension. The contribution of the
hadronic component contains only hadronic
parameters, although it may have a stronger dependence on $N_c$.
Examining the cusp with
an appropriately high resolution
 one would observe that it is smoothed on the
hadronic  scale, where the hadronic component of the
axion wall ``sandwich'' would
become visible.
The cusp carries a finite contribution to the
wall tension which cannot be calculated in the
 low-energy approximation \cite{KoganKovnerShifman}.
To this end one needs to consider the hadronic core
explicitly.
The tension of the core $T_{\rm core} \sim \Lambda^3 N_c$,
while the tension of the axion
halo $T_{\rm halo} \sim f\Lambda^2$ (in pure gluodynamics).

We pause here to make a comment on the literature.
The consideration of the axion walls in conjunction with
 hadrons dates back to
the work of Huang and Sikivie, see Ref. \cite{Sikivie:1982qv}.
This work treats the Weinberg-Wilczek $N=2$ axion in
QCD with two light flavors, which  is replaced by a chiral Lagrangian
for the pions, to the leading order (quadratic in derivatives and
linear in the light quark masses). It is well-known
\cite{Witten:1980sp,DiVecchia:1980ve,Smilga:1999dh} that
in this theory the crossover phenomenon takes place  at
$m_u=m_d$. In the realistic situation,
$(m_d-m_u)/(m_d+m_u)\sim 0.3$ considered in Ref.
\cite{Sikivie:1982qv},
there is no crossover. The pions can be integrated over,
leaving one with an effective Lagrangian for the axion of the type
(\ref{dtwo}) (with $\alpha \to 2\alpha$). The potential is not pure
cosine,
 higher harmonics occur too.
The axion halo exhausts the wall, there is no
hadronic core in this case.

At the same time, Huang and
Sikivie (see Ref. \cite{Sikivie:1982qv})
found   an explicit solution for the ``$\pi^0$" component of the
wall. In fact, this is an illusion.
The Huang-Sikivie (HS) solution refers to the
{\em  bare}
$\pi^0$ field. To find the physical $\pi^0$ field one must
diagonalize the mass matrix
at every given value of $\alpha$ (the bare $f\alpha$
is the physical  axion field up to
  small corrections $\sim f^2_{\pi}/f^2$
where $f_\pi$ stands for the pion decay constant).  Once this is done, one
observes that the physical pion field,
which is a combination of the bare pion and $f\alpha$,
is not excited in  the
HS solution. The equation (2.16) in the HS paper  is exactly the condition
of vanishing of the physical pion in the wall profile.
This explains why the wall thickness in the HS work
is of order $m_a^{-1}$,
with no traces of  the $m_\pi^{-1}$ component.
The crossover of the hadronic vacua at
$\alpha = \pi /2$ (remember, this is $N=2$ model)
could  be recovered in the Huang-Sikivie analysis at
$m_u=m_d$. However, the chiral pion Lagrangian
predicts  in this case the vanishing of the pion mass
in the middle of the wall, for accidental reasons. This is explained in
detail by A. Smilga, Ref.  \cite{Smilga:1999dh}.

\subsection{An Illustrative model}

To find the axion walls with D-wall core
one has to  solve QCD, which is way beyond our possibilities.
Our task is more modest. We would like to obtain a qualitative
description of the axion wall sandwich
which, with luck, can become semi-quantitative. To this end we want to
develop toy models.
An obvious requirement to any toy model
is that it must qualitatively reproduce the basic features of
the vacuum structure which we expect in QCD.
In  SUSY gluodynamics
it was possible to write down
a toy model with a ${\bf Z}_{N_c} $ symmetry
\cite {GabadZ} which ``integrates in"
the heavy degrees of freedom and
allows one  to
investigate the BPS domain walls in the
large $N_c$ limit \cite {DGK} (see also
\cite {DK}).  We will
suggest  a similar model
in (nonsupersymmetric) QCD, then switch on axions,
and study the axion domain walls in a semi-realistic setting.
In this model we will  be able to find exact  solutions
for D-walls and axion walls.

Here we suggest a simple toy model which has a proper vacuum
structure.
In what follows  an appropriate (complex)  glue order parameter is
denoted by $\Phi$. The modulus and phase of this field
describe  respectively the $0^{++}$
and $0^{-+}$ channels of the theory.

Our  toy model Lagrangian is
\beq
{\cal L} &=& N_c^2 (\partial_\mu \Phi)^*(\partial_\mu \Phi)
- V(\Phi, \Phi^* )\, , \qquad  V= V_0 + V_1\, ,\nonumber \\[0.2cm]
V_0 &=&  N_c^2 A^2 \left| 1-\Phi^{N_c}e^{-i\theta}\right|^2\,, \nonumber
\\[0.2cm] V_1 &=& \left\{- {{\cal X} N_c ^2\over 2}\,  \Phi \left [ 1+{1\over
N_c }
(1-
 \Phi^{N_c} e^{-i\theta})  \right] +{{\cal X} N_c ^2\over 2}\right\}
+\mbox{H.c.} \,.
\label{Phi}
\eeq
Here $A$ is a numerical constant of order one,
and ${\cal X}$ is the  vacuum topological susceptibility
in pure gluodynamics (note that ${\cal X}$
is  independent of $N_c $).  The scale parameter $\Lambda$ is
set to unity.

This model  has the vacuum family composed of
$N_c$ states.  Indeed, the minima of the energy are
determined from the equations
\beq
\left. {\partial V \over \partial \Phi }\right|_{\rm vac}=
\left. {\partial V \over \partial { \Phi^*} }\right |_{\rm vac}=0\,,
\eeq
which have the following solutions (we put temporarily $\Lambda=1$):
\beq
\Phi_{\ell \rm vac}={\rm exp} \left ( i\, {\theta+2\pi \ell \over N_c }
\right )\,,
\quad \ell = 0,1,..., N_c-1\,.
\label{vsol}
\eeq
In the $\ell$-th minimum $V_0$ vanishes, while $V_1$
produces a non-vanishing vacuum energy density,
\beq
{\cal E}_\ell = {\cal X} N_c ^2\left\{
1-\cos\left( {\theta+2\pi \ell \over N_c }\right)\right\} \,.
\eeq
For each given $\theta$ the genuine vacuum is found
by minimization,
\beq
{\cal E} (\theta ) = N_c ^2 {\cal X}  \, \mbox{min}_\ell
\left\{
1-\cos\left( {\theta+2\pi \ell \over N_c }\right)
\right\}
\,.
\label{cos}
\eeq
The remaining $N_c-1$ minima are quasivacua.
Once the heavy field
$\Phi$ is integrated out, the vacuum energy is
given by the expression (\ref {cos}); it has cusps at
$\theta =\pi , 3\pi$ and so on.
Needless to say that  the
potential (\ref {Phi}) has no cusps.

We will first consider the model (\ref{Phi}) without the axion field,
at $\theta = 0$, in the limit $N_c\to\infty$. In this limit
the false vacua from the vacuum family  are stable.

The classical  equation of motion
defining the wall is
\beq
N_c ^2\, {\Phi^*}^{\prime\prime}={\partial V\over \partial \Phi}\,,
\label{eq}
\eeq
where primes denote differentiation with respect to $z$
(we  look for a solution which depends
on the $z$ coordinate only).

This is  a differential equation of the  second order.
It is possible, however, to reduce it to a
first order equation. Indeed, Eq. (\ref{eq})
has an obvious ``integral of motion" (``energy"),
\beq
N_c ^2 ~{ \Phi^*}^\prime \Phi^\prime - V = \mbox{Const} = 0\,,
\label{V}
\eeq
where the second equality follows from the boundary conditions.
In the large $N_c $ limit one can  parametrize the field
$\Phi$ as follows ($\rho \sim 1$):
\beq
\Phi\equiv 1+ {\rho\over N_c }\,.
\label{Phirho}
\eeq
Taking the square root of Eq.
(\ref {V}), substituting Eq.
(\ref{Phirho}) and neglecting the terms  of the subleading order in
$1/N_c
$ we arrive at
\beq
{\bar \rho}^\prime =i A N_c \, \left  (1-{\rm exp} \rho\right )\,.
\label{BPS}
\eeq
The phase on the right-hand side can be chosen arbitrarily.
The choice in Eq. (\ref{BPS})
is made in such a way as to make it compatible with the boundary
conditions
for the wall interpolating between $\Phi_{\rm vac} = 1$ and
$\Phi_{\rm vac} = \exp (2\pi i / N_c )$.
This is precisely the expression that defines
the domain walls in SUSY \g~ \cite {DGK,DK}. It is not surprising
that the same equation determines the D-walls in non-SUSY \g~--
the fermion-induced  effects are not important for
D-walls in the large $N_c $ limit.

The solution of this equation was obtained in \cite {DK}.
In the parametrization $\rho=\sigma+i\tau$
the solution  takes the form:
\beq
&&\cos \tau =(\sigma +1)~{\rm exp}(-\sigma), \nonumber \\[0.3cm]
&&\int_{\sigma(0)}^{\sigma(z)}[{\exp}(2t)-(1+t)^2]^{-1/2}~dt=-
A  N_c |z|\, .
\label{Dsolution}
\eeq
The real part of $\rho$  is   a bell-shaped function with
 an extremum at zero; it vanishes  at $\pm \infty$.
The imaginary part of $\rho$, on the other hand, changes its value from
$0$ to
$2\pi$.  This determines a D-wall  in the large $N_c $ gluodynamics.
The width of the wall scales as $1/N_c $.

The solution presented above is exactly the same as in SUSY
gluodynamics.
This is not surprising since the
{\em ansatz} (\ref{Phirho}) implies that $V_1$
does not affect the solution --
its impact is subleading in $1/N_c$, while
$V_0$ is exactly the same as in the  SUSY-gluodynamics-inspired model
of Ref. \cite{DGK}. Moreover, for the same reason
the domain wall junctions emerging in this model
will be exactly the same as in  the  SUSY-gluodynamics-inspired model
\cite{GabadShifman}. Inclusion of $V_1$
in the subleading order makes the wall to decay.

Inclusion of the $N=1$  axion field amounts
to the replacement
$$
\theta \to \theta +\alpha
$$
in Eq. (\ref{Phi}), plus
the axion kinetic term
\beq
{\cal L}_{\rm kin} = \left ({f^2~+~2\Phi^*\Phi \over 2}\right )
(\partial_\mu \alpha )^2
+ iN_c (\partial_\mu \alpha ) (\Phi^* \partial_\mu \Phi -
\Phi \partial_\mu \Phi^*)\,.
\eeq
The occurrence of the mixing between $\alpha$ and the phase of $\Phi$
is necessary, as is readily seen from the softly broken SUSY
 gluodynamics. (To get the potential of the type (\ref{Phi})
in this model, one must eliminate the $G\tilde G$ term by a chiral
rotation.
Then $m \to m\exp ((\theta +\alpha)/N_c)$ and, additionally
one gets $\partial_\mu\alpha\times$ [the gluino axial current].)
The term $2\Phi^*\Phi $ in the brackets has to be included
to reproduce the correct mass for the axion after the
physical heavy  state is integrated out.
The presence of this term signals  that QCD
dynamics generates not only the potential for the axion but
also modifies its kinetic term. On the other hand, since
$\Phi^*\Phi\le\Lambda^2$ and, moreover, $\Lambda<<f$, this
term can be neglected for all practical purposes.

We are interested in the configuration with
$\alpha$ interpolating between 0 and $2\pi$.
The phase of $\Phi$ will first adiabatically  follow $\alpha/N_c$,
then at $\alpha \approx\pi$, when the phase of $\Phi$
is close to $\pi /N_c$, it will very quickly
jump by $-2\pi/N_c$, and then it will continue
to grow  as $\alpha /N_c$, so that when $\alpha$ reaches $2\pi$
the phase of $\Phi$ returns to zero. This jump is continuous, although
it occurs at
a scale much shorter than $m_a^{-1}$.  This imitates
the D-wall core of the axion wall.
One cannot avoid forming this core, since otherwise
the interpolation would not connect degenerate states --
on one side of the wall we would  have (hadronic) vacuum, on the other
side an excited state.

In the  large $N_c$ limit one can be somewhat more quantitative.
Indeed, in this approximation the model admits the exact solutions.
The gluonic core of the wall has the same form as before, Eq.
(\ref {Dsolution}), but the phase $\tau$ is now substituted by
the superposition  $\tau -(\alpha +\theta )$  since  the
axion field is mixed with the phase of the $\Phi$ field.

This very narrow core is surrounded by a diffused axion halo.
The axion field is described in this halo
by the solution to the Lagrangian
(\ref {axYM}). This takes the form:
\beq
\theta~+~\alpha (z)~=~-2\pi~+~4N_c~{\rm arctan}\left
(e^{m_az}~{\rm tan}{\pi\over 4N_c}
\right )~,~~~~z<0, \nonumber \\
\theta~+~\alpha (z)~=~-
~4N_c~{\rm arctan}\left (e^{-m_az}~{\rm tan}{\pi\over 4N_c}
\right )~,~~~~z>0~.
\label{axionwalls}
\eeq
Thus, we find explicitly the stable axion wall with a D-wall core.
Note that this is a usual ``$2\pi$'' wall
as  it separates two identical hadronic vacua.
As we discussed in the introduction, this wall can decay quantum
mechanically.
However, its lifetime is infinite for all practical purposes.
Moreover, this wall  is harmless cosmologically.  It  will   be
produced bounded by global axion strings in the early universe.
Bounded  walls shrink  very quickly
by decaying  into axions and hadrons.

\section {Axions and vacuum structure in QCD with light quarks}

So far we discussed pure gluodynamics with the axion.
Our final goal is to study QCD  with $N_f= 3$.
There are two, physically distinct regimes to be considered in this
case. In real QCD
\beq
m_u,~m_d~\ll~m_s~\sim~{\Lambda\over N_c}~,
~~~m_u,~m_d,~m_s~\ll~\Lambda~.
\label{I}
\eeq
In this regime the consideration of the chiral Lagrangians \cite
{Witten:1980sp,DiVecchia:1980ve,Smilga:1999dh}, does not exhibit the
vacuum family. We will comment on why
the light quarks screen the vacuum family of the  glue sector,
so that the axion domain wall provides no access to it.
In the limit (\ref{I}) the effects due to the
D-walls will be marginal.

On the other hand, in the genuinely  large $N_c$ limit
\beq
{\Lambda\over N_c}~\ll~m_u,~m_d~\ll~m_s~\ll~\Lambda,
\label{II}
\eeq
physics is rather similar to that of pure gluodynamics. The light quarks
are too heavy to screen the vacuum family of the  glue sector

In what follows we study the axion walls
and their hadronic components in the  limits
(\ref {I}) and (\ref {II}), separately.

\subsection{ One light quark}

To warm up, let us start from the theory with one light quark.
In the limit of large $N_c$ this introduces a light meson, ``$\eta '$".
An appropriate effective Lagrangian can be obtained by combining
the vacuum energy density of gluodynamics with what remains from
the chiral
Lagrangian at $N_f =1$,
\beq
{\cal L} &=&
 \frac{ F^2}{2} (\partial_\mu \beta)^2
-V(\beta )\,,\nonumber\\[0.3cm]
V &=& -m_q\Lambda^3N_c \cos\beta +
 \min_\ell \left\{- N_c^2\Lambda^4
\cos \frac{\beta +\theta +2\pi \ell}{N_c}
\right\}\, .
\label{dsix}
\eeq
Here $\beta$ is the phase of $U \sim \bar q_L q_R$, while  $F^2 \sim
\Lambda^2 N_c$ is the ``$\eta '$'' coupling constant squared. The product
$F\beta$ is the ``$\eta '$'' field.
The first term in $V$ corresponds to the quark mass term,
${\cal M}U$ + h.c.
At $N_c =\infty$ the second term in $V$ becomes
$(\beta +\theta )^2$. It corresponds to $(i\mbox{ln det} U +\theta )^2$
in Eq. (11) in  \cite{Witten:1980sp}. The subleading in $1/N_c$ terms
sum up into a $2\pi$ periodic function of the cosine type, with the cusps.
It is unimportant that we used cosine in Eq.  (\ref{dsix}).
Any $2\pi$ periodic function of this type would lead to the same conclusions.
The second term in Eq.  (\ref{dsix}) differs from the vacuum energy
density in gluodynamics by the replacement $\theta \to \beta +\theta$.

If $
m_q \ll{\Lambda}/{N_c}$,  the first term in $V$ is
a small perturbation; therefore,
in the vacuum, $\beta +\theta =2\pi k$, and, hence,
the $\theta$ dependence of the vacuum energy is
\beq
{\cal E}_{\rm vac} (\theta ) = -m_q\Lambda^3N_c \cos\theta\,.
\eeq
It is smooth,  $2\pi$ periodic and proportional to $m_q$
as it should be on general grounds in the theory with one light quark.

The condition $
m_q \ll{\Lambda}/{N_c}$ precludes us from sending
$N_c\to \infty$.
The would be
``$2\pi$" wall in the variable $\beta$ is expected to be
unstable.
 This is due to
the fact that at $N_c \sim 3$ the absolute value of the quark condensate
$\bar \psi \psi$  is not  ``harder" than the phase of the condensate
$\beta$, and the barrier preventing the creation of holes in the
``$2\pi$" wall is practically absent.

If one closes one's eyes on this instability
one can estimate that the
 tension of the ``$\eta '$" wall is proportional to $\Lambda^3
N_c^{1/2}$,
with a small correction $m_q\Lambda^2 N_c^{3/2}$ from the quark mass
term.
The tension of the D wall core is, as previously, $\Lambda^3N_c$.

In the opposite limit
\beq
m_q \gg\frac{\Lambda}{N_c},\qquad
\mbox{but $m_q$  still  $ \ll\Lambda$}\, ,
\label{limtwo}
\eeq
the situation is trickier. Now the first term in $V$ is dominant, while the
second is a small perturbation.
 There are $N_c$ distinct vacua in the theory,
\beq
\beta_{\ell}~=~-{2\Lambda \over m_q N_c}~(\t+2\pi\ell)~.
\eeq
Then the $\theta$ dependence of the vacuum energy density
is
\beq
{\cal E}_{\rm vac} (\theta ) = \Lambda^4 \min_\ell ( \theta +2\pi\ell
)^2\,,
\label{en2}
\eeq
this is  similar to that in the theory without  light quarks (i.e.,
the same as in
gluodynamics).
 The ``$\eta '$"   wall is stable at $N_c\to\infty$, with a
a D-wall core in its center. The $\eta'$ wall is a ``$2\pi$''
wall.

From this standpoint the quark with the mass
(\ref{limtwo}) is already heavy, although
the ``$\eta '$" is still light on the scale of $\Lambda$,
$$
M_{\eta '}\sim m_q^{1/2}\Lambda^{1/2} \ll \Lambda\,.
$$

\vspace{0.2cm}

So far the axion was switched off. What changes if one
includes it in the theory?

The Lagrangian now becomes
\beq
{\cal L} &=&
 \frac{ F^2}{2} (\partial_\mu \beta)^2  +  \frac{ f^2}{2} (\partial_\mu
\alpha)^2
-V(\beta ,\alpha )\,,\nonumber\\[0.3cm]
V &=& -m_q\Lambda^3N_c \cos\beta +
 \min_\ell \left\{- N_c^2\Lambda^4
\cos \frac{\beta +\alpha +2\pi \ell}{N_c}
\right\}\, ,
\label{dseven}
\eeq
where the $\theta$ angle is absorbed in the definition of the
axion field.

The bare ``$\eta '$" mixes with the bare axion. It is easy to see that
in the limit $
m_q \ll{\Lambda}/{N_c}$
the physical ``$\eta '$" is proportional to $\beta + \alpha$, rather than to
$\beta$. Therefore, even if we force the axion wall to develop,
(i.e. $\alpha$ to evolve from $0$ to $2\pi$)
the ``$\eta '$" wall need not develop. It is energetically expedient to
have $\beta + \alpha=0$.  Thus,  the
effect of  the axion field on the hadronic
sector is totally screened by a dynamical phase $\beta$
coming from the quark condensate. In other words,
the axion wall with the lowest tension
corresponds to the frozen physical ``$\eta '$",
 $$\beta + \alpha = 0\,.$$
There is no hadronic core.  The tension of this wall is determined from the
term
$\propto m_q\Lambda^3N_c$.

(If one wishes, one could add the ``$\eta '$" wall to the axion wall.
Then the ``$\eta '$" wall, with the D-wall core will appear
 in the middle of the
axion wall, but they are basically unrelated.
 This will be a secondary
phenomenon,
and the D wall core will be, in fact, the core of the
``$\eta '$" wall rather than the axion wall.)

If the quark mass is such that (\ref{limtwo})
applies, then
the axion field $\alpha$ cannot be screened,
since we cannot freeze $\beta +\alpha$
everywhere in the axion wall profile at zero -- at $m_q\gg\Lambda/N_c$,
$\beta$ is proportional to the physical ``$\eta '$" and is much heavier
than
the axion field.
Thus,  in this case the axion wall
will be described by the Lagrangian (\ref {axYM}) and will have a D-wall
core.
One may also add, on top of it, the
``$\eta '$" wall.
 This will cost
$m_q^{1/2}\Lambda^{5/2} N_c$ in the wall tension --
still much less than $\Lambda^3 N_c$ of the D-wall
core of the axion wall.

The limit (\ref{limtwo}) is unrealistic. Moreover, in this limit
the D walls taken in isolation, without the
axion walls, are stable by themselves, although they
interpolate between nondegenerate states \cite{ShifmanM}.

\subsection{Three Light Quarks}

Let us turn the case of three light flavors.
The physical picture is quite  similar to that of
the one-flavor case, see Sec. 7.1.

We assume  the mass matrix ${\cal M}$  in the meson Lagrangian
to be  diagonal. Therefore, we
will  looking for a diagonal
$U(3)$ meson matrix which minimizes the potential,
\beq
U~=~{\rm diag}~\left ( e^{i\phi_1},~~e^{i\phi_2},~~e^{i\phi_3}~\right )~.
\label{U}
\eeq
The potential takes the form
\beq
V ~=~ -\sum_i m_i \Lambda^3 N_c {\rm cos} \phi_i~ +~
 \min_\ell \left \{- N_c^2 \Lambda^4
\cos ~\frac{\sum_i \phi_i +\theta +2\pi \ell}{N_c}
\right\}~.
\label{V3}
\eeq
As before, we will consider two limiting cases, (\ref {I}) and (\ref {II}).

\vspace{0.2cm}

Let us switch off the axion field first.
In the limit of genuinely light quarks, Eq. (\ref {I}),
when the second term in the potential (\ref{V3})
is dominant,
 the  solutions
for
$\phi$'s were found in \cite {Witten:1980sp,DiVecchia:1980ve}.
They satisfy  to the relation
$\phi_3\simeq 0$ and $\phi_1+\phi_2 =-\t$.
The corresponding expression for the vacuum energy density is
\beq
{\cal E}_{\rm vac}(\t) ~=~-
N_c~\Lambda^3~\sqrt{~m_u^2+m_d^2+2m_u~m_d\cos \t}~.
\eeq
As in Sec. 7.1, we deal here with a smooth single-valued function of $\theta$.
The inclusion of the axion replaces $\theta\to\theta +\alpha\to\alpha$.
The physical $\eta'$ field is
given by the sum $\sum_i\phi_i+\alpha$.
It is energetically  favorable to freeze this state.
Thus, the situation is identical to that in the one-flavor case:
even if the axion wall is forced to develop, the physical $\eta'$
wall (which is now the $\sum_i\phi_i+\alpha$ wall) does not have
to occur.  Effectively, the vacuum angle is screened,
and there is
no D-wall core in the axion wall.

If, nonetheless, the $\eta '$  wall is
formed  due to  some cosmological initial conditions, it will have a D-wall
core (albeit the $\eta '$  wall is unstable in the limit at hand
and cannot be considered in the static approximation).  The
would-be $\eta '$  wall is independent of the axion wall;
 its effect on the
axion wall  formation is rather irrelevant.

In addition to this, a ``$2\pi$''
wall could  develop for nonsinglet mesons at certain values of the quark
masses. There is nothing new we could add to this issue
which is decoupled from the issue of the vacuum family in the glue sector
and D-walls.

We now pass to the opposite limit  (\ref {II}), when the first term in
 the potential (\ref{V3}) is dominant.
As in Sec. 7.1,
there are $N_c$ distinct vacua with the energy given by (\ref {en2}).
It is straightforward to show that the
potential for the axion in this case is of the form (\ref {axYM}),
with the cusps which signal the presence of the  D-wall core.
This is similar to what happens in  gluodynamics.
One cannot avoid having an  $\eta '$ wall in the middle of the axion wall,
which entails a D-wall too.
The D-walls  separate the  degenerate vacua.
Since they ``live" in the middle of the axion wall, they are perfectly stable.

(In addition, there can be
``$2\pi$'' walls in either
of $\phi$'s or their linear combinations. However, these latter are
unstable and do not appear in the physical spectrum of
the theory.)
\vspace{0.5cm} \\
{\bf Acknowledgments}
\vspace{0.1cm} \\

The authors would like to thank S. Dimopoulos, A. Smilga,
L. Susskind and E. Zhitnitsky for usefull discussions.
The work is supported by  DOE grant DE-FG02-94ER408.

\end{document}